\documentclass{natureprintstyle}

\usepackage{graphicx,float}
\usepackage{epstopdf,caption,longtable}
\usepackage{float} 
\usepackage{color} 
\usepackage{amsmath}
\usepackage{amssymb,xspace,array,chngcntr}
\usepackage{soul}

\newcommand{  \Hbeta    }{\ifmmode {\rm H}\beta \else H$\beta$\fi}
\newcommand{\gadget}{{\sc GADGET-3}}
\newcommand{\sunrise}{{\sc  SUNRISE}}
\newcommand{\sextractor}{{\sc SExtractor}}
\newcommand{  \NevIR    }{\ifmmode {\rm Ne}\,\textsc{v}\,14.3\,\mu {\rm m} \else Ne\,\textsc{v}\,$14.3\,\mu$m\fi}

\newcommand{\lumagn}{55}
\newcommand{\lumagnlate}{1}
\newcommand{\lumagnper}{1.8\%}
\newcommand{\lowlumagn}{73}
\newcommand{\lowlumagnlate}{2}
\newcommand{\lowlumagnper}{2.7\%}
\newcommand{\inactive}{176}

\newcommand{\inactivelate}{2}

\def \hst{{\it HST\ }}
\def \hstsh{{\it HST}}

\title{A population of luminous accreting black holes with hidden mergers}

\author{Michael J. Koss$^{1,2}$, Laura Blecha$^3$, Phillip Bernhard$^2$, Chao-Ling Hung$^4$, Jessica Lu$^5$, Benny Trakhtenbrot$^{6,7}$, Ezequiel Treister$^8$, Anna Weigel$^2$, Lia F. Sartori$^2$, Richard Mushotzky$^9$, Kevin Schawinski$^2$, Claudio Ricci$^{10,11,12}$,  Sylvain Veilleux$^9$, and David B. Sanders$^{13}$}

\def\micron{{\mbox{$\mu{\rm m}$}}}
\def\arcsec{{\mbox{$^{\prime \prime}$}}}

\def\arcsec{{\mbox{$^{\prime \prime}$}}}

\def\erg{{\rm\thinspace erg}}

\def\Msun{\hbox{$\rm\thinspace M_{\odot}$}}

\def\s{{\rm\thinspace s}}

\def\ergps{\hbox{$\erg\s^{-1}\,$}}

\def\apjl{\hbox{ApJL}}
\def\apjs{\hbox{ApJS}}

\def\aap{\hbox{AAP}}
\def\micron{{\mbox{$\mu{\rm m}$}}}
\def\arcsec{{\mbox{$^{\prime \prime}$}}}

\newcommand{\OIII}{\ifmmode \left[{\rm O}\,\textsc{iii}\right]\,\lambda5007 \else [O\,{\sc iii}]\,$\lambda5007$\fi}

\def\H{{\it HST}}

\def \swiftbat {{\it Swift}/BAT\ }

\def \swiftbatsh {{\it Swift}/BAT}

\begin{document}

\maketitle

\begin{affiliations}
\item Eureka Scientific Inc, Oakland, CA, USA
\item Institute for Particle Physics and Astrophysics, ETH Z\"{u}rich, Z\"{u}rich, Switzerland
\item Department of Physics, University of Florida, Gainesville, FL, USA
\item Department of Physics, Manhattan College, New York, NY, USA
\item Department of Astronomy, University of California, Berkeley, CA, USA
\item Department of Physics, ETH Z\"{u}rich, Z\"{u}rich, Switzerland
\item School of Physics and Astronomy, Tel Aviv University, Tel Aviv, Israel 
\item Instituto de Astrof\'{\i}sica, Facultad de F\'{\i}sica, Pontificia Universidad Cat\'{o}lica de Chile, Santiago, Chile
\item Department of Astronomy and Joint Space-Science Institute, University of Maryland, College Park, MD, USA
\item N\'{u}cleo de Astronomía de la Facultad de Ingenier\'{\i}a Universidad Diego Portales, Santiago, Chile
\item Kavli Institute for Astronomy and Astrophysics, Peking University, Beijing, China
\item Chinese Academy of Sciences South America Center for Astronomy, Santiago, Chile
\item Institute for Astronomy, University of Hawaii, 2680 Woodlawn Drive, Honolulu, HI 96822, USA

\end{affiliations}
\begin{abstract}
Major galaxy mergers are thought to play an important part in fuelling the growth of supermassive black-holes\cite{DiMatteo:2005:604}.  However, observational support for this hypothesis is mixed, with some studies showing a correlation between merging galaxies and luminous quasars\cite{Goulding:2018:S37,Donley:2018:63} and other studies showing no such association\cite{Villforth:2017:812, Chang:2017:19}.  Recent observations have shown that a black hole is likely to become heavily obscured behind merger-driven gas and dust, even in the early stages of the merger, when the galaxies are well separated\cite{Kocevski:2015:104,Koss:2016:85,Ricci:2017:stx173} (5 to 40 kiloparsecs). Merger simulations further suggest that such obscuration and black-hole accretion peaks in the final merger stage, when the two galactic nuclei are closely separated\cite{Hopkins:2005:L71} (less than 3 kiloparsecs).  Resolving this final stage requires a combination of high-spatial-resolution infrared imaging and high-sensitivity hard- X-ray observations to detect highly obscured sources. However, large numbers of obscured luminous accreting supermassive black holes have been recently detected nearby (distances below 250 megaparsecs) in X-ray observations\cite{Baumgartner:2013:19}.   Here we report high-resolution infrared
observations of hard-X-ray-selected black holes and the discovery of
obscured nuclear mergers, the parent populations of supermassive black-hole mergers. We find that luminous obscured black holes (bolometric luminosity higher than $2\times10^{44}$ ergs per second) show a significant ($P<0.001$) excess of late-stage nuclear mergers (17.6 per cent) compared to a sample of inactive galaxies with matching stellar masses and star formation rates (1.1 per cent), in agreement with theoretical predictions. Using hydrodynamic simulations, we confirm that the excess of nuclear mergers is indeed strongest for gas-rich major-merger hosts of obscured luminous black holes in this final stage.  As the parent population of supermassive black hole mergers, the study of nuclear mergers can provide crucial benchmarks for models of black-hole inspiral and gravitational-wave signals.


\end{abstract}

The Burst Alert Telescope (BAT) on the Neil Gehrels $Swift$ Observatory has surveyed the entire sky at unprecedented depths in the ultra-hard X-ray band (14--195\,keV) and primarily detects accretion on to supermassive black holes (SMBHs) at the centres of nearby galaxies.  Detection in the ultra-hard X-ray band is possible even when obscuring gas and dust in the host galaxy significantly attenuates the ultraviolet, optical, and/or softer X-ray emission around the growing black holes.  At the distance to the nearest luminous AGN (about 220 Mpc or $z{\approx}0.05$) ground-based optical imaging typically achieves a resolution of order 1\arcsec\ or 1~kpc in the host galaxy.  This spatial resolution is not sufficient to resolve the final merger stage in the host galaxies down to the 100s of pc scales. However these can be resolved with near-infrared adaptive optics, which provide an
improvement by a factor of 10 in spatial resolution (about 0.1\arcsec).  These scales are still above the black hole sphere of influence, which is of the order 10-100 pc for black holes with masses of $10^7-10^9$ \Msun\ (\Msun, solar mass).


We observed 96 nearby ($z{<}0.075$) black holes detected by \swiftbat\
in the hard X-ray band. The black holes were selected at random over
a wide range of luminosities using the adaptive optics system on the Keck 2 telescope at the W. M. Keck Observatory with a near-infrared
camera (NIRC2). These near-infrared observations (2.1 \micron) include both obscured and unobscured sources.   Our observations have an average spatial resolution of $0.13\arcsec$, about a factor of 10 better than previous ground-based surveys.    We combined these adaptive-optics observations
with available high-resolution archival Hubble Space Telescope
(\hstsh) near-infrared images of 64 \swiftbatsh-detected active galactic
nuclei (AGN) with an average spatial resolution of $0.17\arcsec$.  These observations
provide the first evidence of a sizeable population of double
nuclei with very small separations (0.3-3 kpc) in late-stage mergers,
which could not be detected in lower-resolution ground-based optical
observations and were not detected in previous near-infrared samples
of AGN observed with the \hst\ (Fig. 1d-f).
     
We separated our sample into obscured and unobscured accreting black holes based on the presence of broad $\Hbeta$ lines in optical spectroscopy from past studies\cite{Koss:2017:74} and into low- and high-luminosity ($L_{\rm bolometric}$ below or above $2{\times}10^{44}$ erg s$^{-1}$, respectively) using X-ray emission\cite{Ricci:2017:17}. We also compared our sample
with 176 inactive galaxies matched in stellar masses and star formation
rates that have high-resolution \hst near-infrared images. Example
high-resolution images of the inactive-galaxy sample are provided in
Extended Data Fig.~3. A comparison of the stellar masses (Extended
Data Fig.~4a), star formation rates (Extended Data Fig.~4b), physical
resolutions (Extended Data Fig.~4c) and the consistency of the control
sample with random inactive galaxies taken from the Sloan Digital
Sky Survey (Extended Data Fig.~6) show that (Fig.~2a) the obscured
luminous black holes show a significantly show a significantly higher fraction ($P<0.001$) of nuclear mergers ($<$3~kpc) than inactive galaxies (\inactivelate/\inactive), luminous unobscured black holes (\lumagnper, \lumagnlate/\lumagn), and low-luminosity black holes (\lowlumagnper, \lowlumagnlate/\lowlumagn).  When comparing the fractions of
nuclear mergers for obscured and unobscured luminous black holes,
the difference is also significant ($P\approx 0.01$). Finally,  a higher proportion of late-stage  mergers ($R<3$ kpc) in luminous obscured black holes is also found when compared to the lower luminosity black holes ($p$-value$\approx$1\%).  At larger separations (3-10 kpc) the fraction of mergers in luminous obscured black holes is higher than the other samples, but this is not statistically significant ($P>0.29$).  All the mergers identified at $R<10$ kpc are listed in Extended Data Table 1.  Finally, we note that even observations in the near-infrared band may sometimes miss nuclear mergers with very heavy extinction\cite{Ohyama:2015:162}, so these measurements should be seen as a lower limit.  
     
     
     While past work has found some nuclear mergers, our study is the
first, to our knowledge, to demonstrate a significant excess ($P<0.001$) of nuclear mergers in luminous obscured AGN as compared to a matched sample of inactive galaxies.  Past studies have typically focused on subsets of AGN galaxies.  For instance, some nuclear mergers have been identified in sources with double peaked \OIII\ emission lines, resulting from the emission from both nuclei\cite{Barrows:2017:129}.  In a sample of 60 double peaked sources observed with NIRC2\cite{Fu:2011:103} only 4/60 (or 6.7\%) were in major mergers with $<$3 kpc separations--a much smaller proportion than that seen in the obscured luminous black holes studied here. Some nuclear mergers have also been detected in the host galaxies\cite{Haan:2011:100} of luminous infrared galaxies with accreting black holes, which have very high star formation rates, probably associated with the merger. However, only one of the ten nuclear mergers in our hard-X-ray sample
is associated with a luminous infrared galaxy (NGC 6240), and none show double peaked \OIII\ emission lines.  This indicates that both of these diagnostics are incomplete indicators of nuclear mergers.

When considering the fractions of galaxies found at various merger
stages, it is critical to consider the corresponding observability timescale,
because the time spent at small separations is thought to be much
shorter than that spent at larger separations. For instance, in a recent
merger simulation study\cite{VanWassenhove:2012:L7}, the time spent at separations $R<3$ kpc can be more than five times shorter than the time spent at separations of 3-10 kpc ($\approx$50 Myr vs. 300 Myr).  Thus, the excess fraction of nuclear mergers in luminous obscured black holes we find in our data is surprising, and it probably reflects a strong link between such mergers and intense black-hole accretion. To compare our observations
with theoretical results more directly, we use a suite of state-of-theart
high-resolution hydrodynamical galaxy merger simulations with the galaxy stellar mass, black-hole mass and black-hole bolometric
luminosity set up to reproduce the accreting black holes and their host
galaxies observed in our study (GADGET-3 code\cite{Springel:2005:1105}; see Methods).  We also consider the simulated mergers at random orientations to the
observer to account for the fact that some mergers would appear closer
simply because of the projection effect (that is, their alignment with the
observer’s line of sight).

     The simulations show that obscured luminous black-hole phases
preferentially occur in the late stages of gas-rich ($M_{\mathrm{gas}}/M_*{<}0.1$), major ($M_1/M_2 {<} 5$) mergers, where $M_{\mathrm{gas}}$ is the gas mass, $M_*$ is the stellar mass and $M_1$ ($M_2$) denotes the galaxy with the larger (smaller) stellar mass. Consistent with our observations, late-stage mergers are less prevalent
in lower-luminosity black holes and inactive galaxies (Fig. 2b).
Finally, our simulations show that obscured luminous black holes,
which occur in the post-merger phase (after the two galactic nuclei
and black holes have merged), contribute as much to the growth of
the obscured black hole as the entire merger phase (R$<$ 30 kpc). We
note that during the late stages of our simulated galaxy mergers, the
black holes spend very little time in an unobscured luminous accreting black-hole phase. These results are consistent with previous theoretical
work\cite{Hopkins:2007:731} that showed that merger-triggered accreting black holes
are preferentially more luminous and obscured than those growing by
stochastic feeding via slower secular processes. This explains the lack of nuclear mergers in low-luminosity AGN seen in a previous large sample
($>$200) high-resolution study of AGN and normal galaxies
using the \hstsh\cite{Hunt:2004:707}.  Moreover, simulations find that although global star formation is enhanced primarily in the early stages of the first merger passage, black-hole growth is minimal until the late merger stages\cite{Capelo:2015:2123} when the galaxies pass within a few kiloparsecs of each other and cause tidal torques that increase nuclear gas inflows.

We simulated a set of mock \hst\ images, targeting redshifted versions of our imaging datasets (see Methods) at the peak of black-hole growth at $z{\approx}1{-}2$ and find that \hst\ would miss the majority of such systems (7/8) at merger separations below 3 kpc owing to insufficient spatial resolution and sensitivity, which is necessary to
identify the nuclear mergers that we find in our low-redshift sample. The upcoming James Webb Space Telescope will provide substantial
improvements in sensitivity. However, the resolution of such nuclear
mergers requires the use of adaptive optics systems in the next generation
of large-diameter ground-based telescopes (for example, the
Thirty Meter Telescope, the European Extremely Large Telescope and
the Giant Magellan Telescope). These will reach resolutions of 300 pc
using adaptive optics at $z{\approx}1{-}2$ --scales that are consistent with the
smallest-separation mergers identified in this study.
      
      With the discovery of gravitational waves emitted from the merger of
stellar-mass black holes, interest in understanding gravitational waves
produced from the merger of supermassive black holes has increased
considerably. The study of nuclear mergers is therefore critical for
comparison with cosmological merger-rate models, because it can help constrain the timescales for supermassive-black-hole inspiral and the
rate of such events, which are likely to be found with gravitational wave
detectors, such as pulsar timing arrays\cite{Verbiest:2016:1267} and the Laser Interferometer Space Antenna\cite{Tang:2018:2249}.     Predictions of the detection rates for these instruments
are based on parameterizations of the merger rates and the
supermassive-black hole-population\cite{Sesana:2018:42}, but these are highly uncertain
and vary by orders of magnitude\cite{Mayer:2013:244008}.  Gravitational-wave observatories
will also struggle with the localization of the sources, which is possible
only with a resolution of the order of 10 square degrees\cite{Lang:2009:094035}, thus requiring a better characterization of their likely precursors. Thus, the study of
nuclear merger fractions and their correlation with galaxy populations can provide crucial benchmarks for models of black-hole inspiral and
the strength of gravitational-wave signals.


\clearpage	
	 
\begin{figure*}
\centering
\includegraphics[width=15cm]{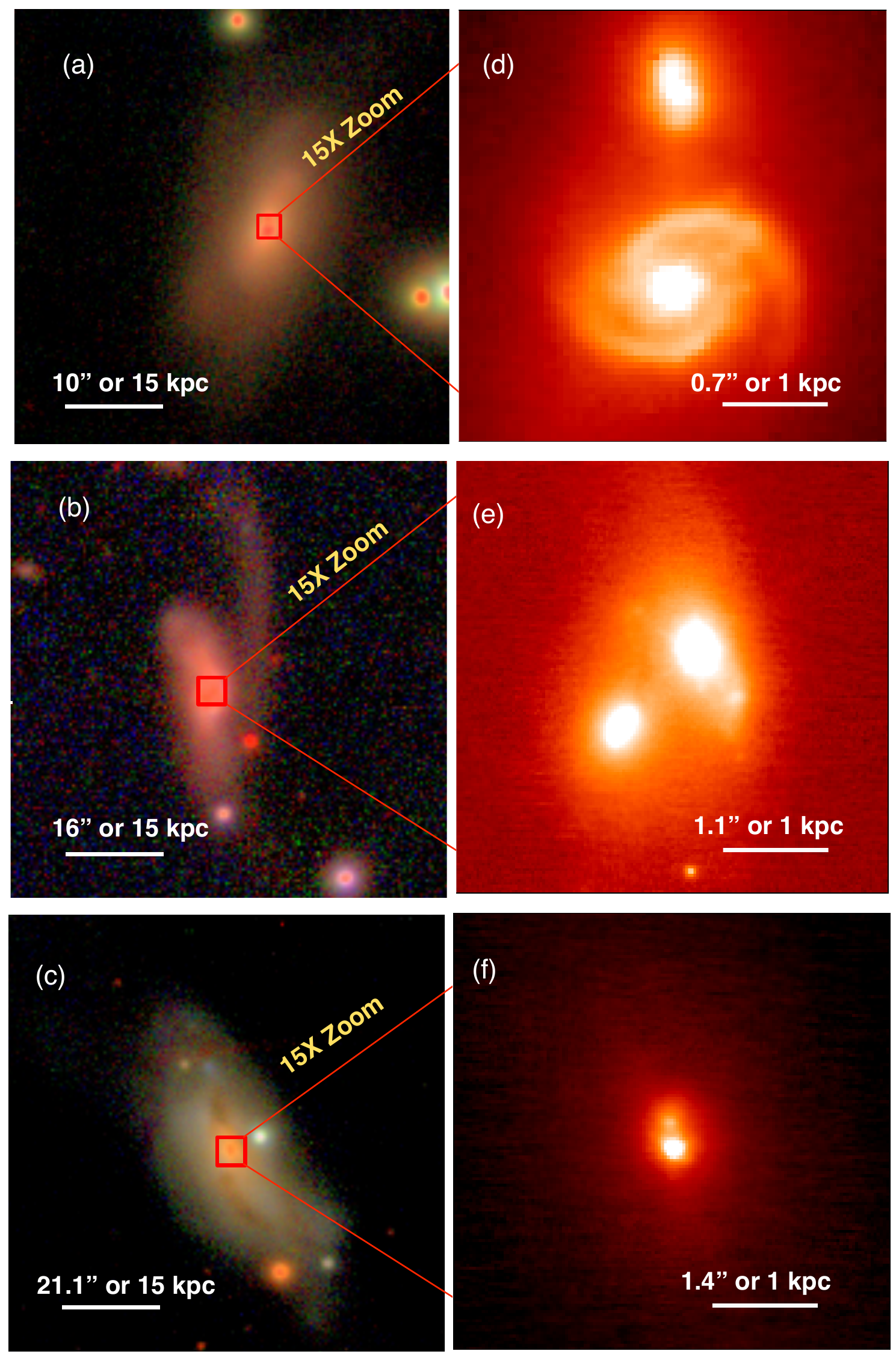}
\caption{\textbf{Example images of final-stage mergers}.     \textbf{a-c,}  Tricolour optical images in the $gri$ bands from the Sloan Digital Sky Survey or the Kitt Peak survey with ~1\arcsec\ angular resolution.  The galaxies shown are 2MASX J01392400+2924067 (\textbf{a}), CGCG 341-006 (\textbf{b}) and MCG+02-21-013 (\textbf{c}). The images are 60 kpc by 60 kpc in size.  Red squares indicate the size of the zoomed-in adaptive optics image on the right.    \textbf{d-f,}  Corresponding near-infrared, $K_{\rm p}$-band (2.12 \micron) adaptive optics images of nuclear mergers taken with the Keck/NIRC2 instrument.  These images are 4 kpc by 4 kpc in size.}
\label{hidden}
\end{figure*}  

\clearpage	
\begin{figure}[H] 
\label{fractions}
\centering
\includegraphics[width=8cm]{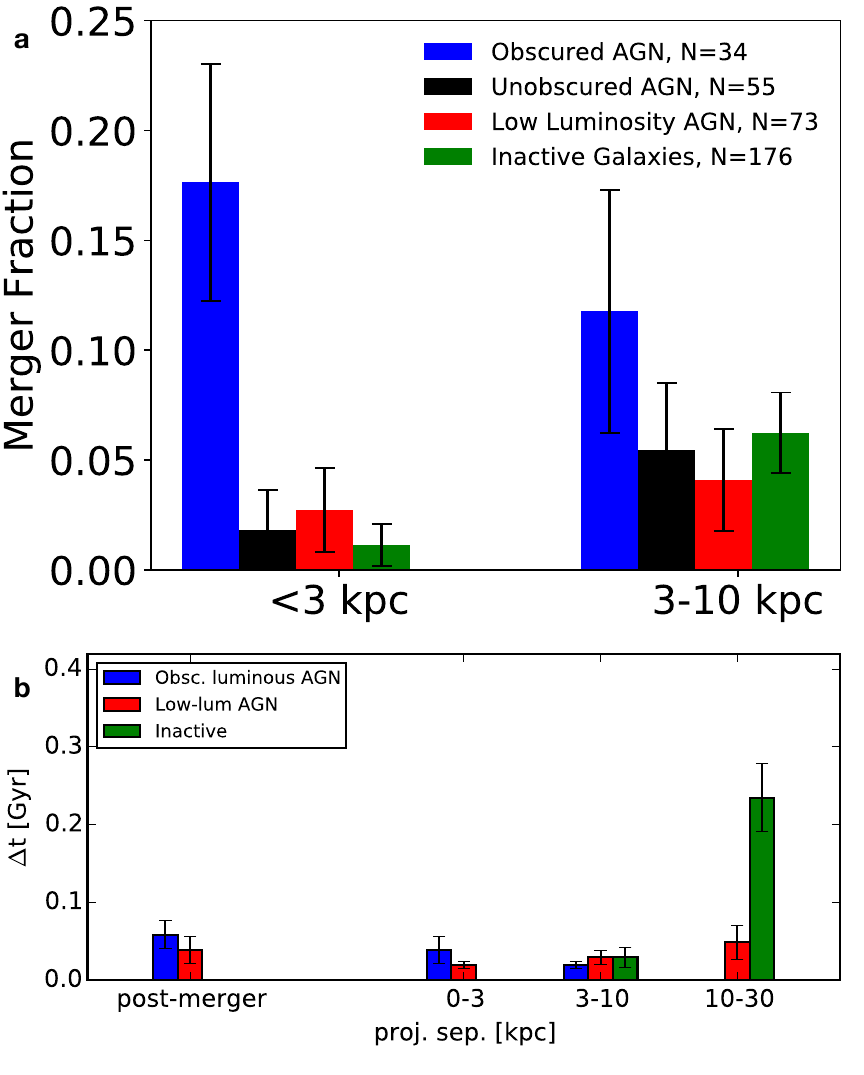}
\caption{\textbf{Fraction of close mergers}.  \textbf{a,}  Fraction of mergers, determined
using high-resolution images obtained with either Keck adaptive optics or
the \H.  
The sample of high-luminosity obscured accreting black holes or
AGN shows a strong excess of small-separation mergers ($<$3 kpc).  Other
types show no significant excess compared to inactive galaxies. Error bars
correspond to 1$\sigma$ confidence intervals. 
\textbf{b,} Results from a suite of gas-rich high-resolution hydrodynamical galaxy merger simulations for a range of viewing angles.  $\Delta t$ is the time spent at a separation range, and error bars represent the median absolute deviation. Our observed merger fractions are consistent with obscured and luminous accreting black holes occurring
primarily in gas-rich major mergers.    }
\end{figure}

\noindent {\bf \huge{Methods}}

\noindent{\bf  Data analysis and sample overview}  We selected our sources from the 70-month {\it Swift}/BAT catalogue which consists a total of 1171 sources, 836 of which are accreting black holes or AGN.  We cross-matched this sample to the Roma Blazar Catalog (BZCAT) catalogue\cite{Massaro:2009:691} to avoid beamed AGN and radio loud AGN which have been extensively studied in past high-resolution studies.    
    
We observed 96 low redshift {\it Swift}/BAT AGN ($0.01<z< 0.075$) that had suitable tip tilt stars with the NIRC2 imager, assisted by an AO system.  The images were taken over nine nights spread between 2012 and 2014.   For bright Seyfert 1 AGN, the nucleus was used as the point-source for tip tilt correction.  Images were taken in the $K_p$ band ($\lambda_{\rm eff}\simeq 2.12$ $\mu$m), and when possible in good seeing conditions, we also took images in the $J$- and $H$- bands (1.25 and 1.63 $\mu$m, respectively). 

We used the wide camera with 40 mas pixel$^{-1}$ and a 40\arcsec\ field of view (FOV). We used a 3 point dither pattern avoiding a known artefact in the lower left part of the field.  For calibration purposes we took dark and flat fields for each night of observation.    
    
The data were reduced using the JLU python code\footnote{https://github.com/jluastro/JLU-python-code/tree/master/jlu}.  The code was modified to ensure that extended galaxy emission features were not subtracted from the background using \sextractor\cite{Bertin:1996:393}.  The images were combined based on weighting by the Strehl ratio of each image.  
	
To increase our sample size we also added 64 BAT-detected AGN which were observed with the \H\ NICMOS or WFC3 cameras. The images were taken in the F105W (1.05 $\mu$m) or F160W (1.60 $\mu$m) filters, with the majority in the F160W band (62/64, 97\%).   Individual frames were co-added, cosmic ray rejected, distortion corrected, and registered using the default values in $\mathtt{Astrodrizzle}$.  For galaxies with $z<0.01$, we use the mean value of redshift independent distance measurements from NASA Extragalactic Database (NED) whenever these are available. We otherwise adopt a cosmology of $\Omega_{\rm m}=0.3$, $\Omega_\Lambda=0.7$, and $H_0=70$\,km\,s$^{-1}$\,Mpc$^{-1}$ for all distances computed. 

Most of the possible galaxy counterparts detected in the images do not have spectroscopy available, so we applied two methods to deal with possible stellar contamination from foreground stars.  Nearby foreground stars and galaxies were identified using segmentation maps produced by \sextractor.  We first applied the stellar classification technique provided by the tool, where a neural network is used as classifier to assign non-stellar objects with 0 and stellar objects with 1.   To separate between galaxies and stars every object with a stellarity index below 0.5 is considered a secondary galaxy.  

Due to the need to rely on tip-tilt stars in some of the AO observations, the AO images typically had more foreground stars than the \H\ ones, which could lead to possible contamination.  We therefore used a second technique to measure the number of stars in the entire FOV that are brighter than our second nearby source divided by the total area searched in the image.  This number is then compared to the search area used to find the nearest companion to provide an estimate for stellar contamination.  We excluded any counterparts where the contamination percentage likelihood was greater than 10\%.  All cases where the contamination percentage likelihood was greater than 10\% also had a stellarity index below 0.5 and were already excluded in agreement with the aforementioned \sextractor stellar classification technique.  

All galaxies classified as extended with low stellar contamination and within 2.5 mag ($\approx$1/10) of the primary AGN or inactive galaxy nucleus were classified as counterparts.


\noindent\textbf{Inactive galaxy control sample.} We developed a large control sample of inactive galaxies by aggregating over 20 years of past \H\ NICMOS and WFC3 surveys conducted in the F160W filter.  For more massive galaxies, which were not well sampled in previous NICMOS surveys, we cross-matched all high-resolution \H\ NIR observations with the NASA-Sloan Atlas\cite{Blanton:2011:31} which includes $\approx$42,000 nearby ($z<0.05$) massive ($M_* > 10^{10}\, M_{\odot}$) inactive galaxies within the footprint of the Sloan Digital Sky Survey. These are typically taken with the \H\ WFC3/NIR camera, due to the small FOV of NICMOS.  We also cross-matched all nearby galaxies ($z<0.05$) from the RC3 catalogue\cite{deVaucouleurs:1995} which covers the entire sky.  Finally, we cross-matched Version 2.1 of the {\it Hubble} Source Catalog, which includes all WFC3/NIR observations with all nearby galaxies ($z<0.05$) from the SIMBAD astronomical database.  To ensure our sample included only inactive galaxies we excluded the 168,941 AGN in the 13th Edition of the Veron-Cetty \& Veron Catalog of Quasars \& AGN\cite{Veron-Cetty:2010:A10}.  We also excluded any galaxies found in clusters, because of the very different environments and generally much higher stellar masses.  Our final control sample includes 385 inactive galaxies, from 37 different \H\ programmes.


If possible we used the Hubble Legacy Archive to download post-processed images.  When these were not available, individual frames were co-added, cosmic ray rejected, distortion corrected in the same way as the \H\ observations of the BAT AGN in our main sample.  The NICMOS images were examined to ensure that the smaller FOV covered the nuclear regions without any significant artefacts after processing.    As with our AGN sample, we used the average of redshift-independent distance measurements from NED, when possible and applied the same method to detect galaxy counterparts.

\noindent{\bf Control sample design.} Although our matching procedure resulted in 385 inactive galaxies observed with \H, many of the relevant \H\ programmes focused on much more nearby and less luminous inactive galaxies than our AGN sample.  This is crucial, as many studies have found that the merger activity and fraction depend on the stellar mass\cite{Patton:9:235,Weigel:2018:2308}.  We therefore measured $H$-band luminosities in both the inactive galaxies and BAT AGN samples, as these are an excellent proxy for the stellar mass with a small scatter of only 0.2 dex\cite{Davies:2015:127}.  For photometry, we use the $H$-band elliptical aperture magnitudes from the 2MASS all sky survey.
    
The merger fraction also depends on star formation rate (SFR).  We therefore used the IRAS 60 \micron\ luminosity as a proxy for SFR.  When the IRAS 60 \micron\ luminosity was measured as an upper limit, we used the 70 \micron\ luminosity from {\it Herschel}/PACS or {\it Spitzer}/MIPS.  We assume a conversion factor of 1.15 between the 70 \micron\ luminosity and 60 \micron\ luminosity based on the average of sources in the sample with both measurements.

The inactive galaxy sample was matched in stellar mass and star formation rate by excluding 117 low stellar mass galaxies ($\langle\log(L_H/L_{\odot})\rangle < 9.7$) and 95 low SFR galaxies ($\log$ ($\nu L_{\nu}$) [erg/s] 60$\mu$m$<43$).  We note that while these 212 inactive galaxies have been excluded from the analysis, we do not find any cases of nuclear ($R<3$ kpc) mergers among them.

The lower luminosity AGN sample has lower average $H$-band luminosity ($\langle\log(L_H/L_{\odot})\rangle = 9.9$) than the inactive control sample ($\langle\log(L_H/L_{\odot})\rangle = 10.1$) or the obscured luminous AGN sample  ($\langle\log(L_H/L_{\odot})\rangle = 10.1$). 
The unobscured luminous AGN sample has slightly higher luminosity ($\langle\log(L_H/L_{\odot})\rangle = 10.2$) than the inactive control sample, however for unobscured AGN, the AGN light can contribute the majority of the emission even in the NIR bands\cite{Koss:2011:57}, so the $H$-band luminosity may overestimate the stellar mass of unobscured luminous AGN.  
As for SFR, the ($\langle\log(\nu L_{\nu}) 60\mu)\rangle = 44.1$) for the inactive galaxies and is the same value for the luminous obscured AGN and luminous AGN. This value is higher than that of the low-luminosity AGN ($\langle\log(\nu L_{\nu}) 60\mu)\rangle = 43.6$).
    
A summary of the different programmes in the control sample based on their abstracts is provided in Extended Data Figure 6.  The majority of the control sample was observed as part of samples studying star formation in LIRGS or large samples of nearby galaxies (70\%, 122/175).  Nearly all of the inactive galaxies images from \H\ were taken in the F160W filter (173/\inactive, 98\%) with the remaining images taken in the F110W filter. The average image resolution was typically FWHM=0.19\arcsec\ for the inactive galaxy sample, slightly lower than the AGN samples which has FWHM=0.12\arcsec.  However, because the nearby inactive galaxies were typically at lower redshift ($\langle z  \rangle = 0.021$) than the AGN sample ($\langle z \rangle =0.034$), the physical scales probed for inactive galaxies ($\langle FWHM \rangle = 79$ pc) were actually smaller than for the AGN sample ($\langle FWHM \rangle = 97$ pc), particularly for the luminous obscured and unobscured AGN ($\langle FWHM \rangle = 134$ pc).    The average Strehl ratio of the AGN sample was 0.45, mainly due to the low Strehl ratios in the AO sample, while the inactive galaxies selected solely from \H\ have a higher Strehl ratio of 0.9.  However,because our study focuses on identifying secondary nuclei that are within 2.5 mag of the bright AGN at the galaxy centers, the reduced sensitivity to very faint objects with AO does not affect our analysis or conclusions.

We also test whether the parent sample of inactive galaxies from which the matched \H\ sample is drawn is itself representative, in a statistical sense, of the nearby galaxies population.    For the nearby galaxies sample we use SDSS DR7 data\cite{Abazajian:2009:543}. We use spectroscopic redshifts from the New York Value-Added Galaxy Catalog (NYU VAGC\cite{Blanton:2005:143}) to limit the sample to the $0.01 < z < 0.05$ range to match the control sample. We extract stellar masses and SFR measurements from the Max Planck Institute for Astrophysics -- John Hopkins University (MPA JHU\cite{Kauffmann:2003:33,Brinchmann:2004:1151}) catalogue which are based on the photometry and emission line modelling to measure stellar masses and SFR values, respectively. We only use sources that are flagged as galaxies in the MPA JHU catalogue ($\approx$90,000).  In order to convert our \hst sample 60 \micron\ emission to SFR we assume standard galaxy templates\cite{Chary:2001:562}. We find that the \H\ sample is representative of the nearby galaxy population in terms of SFR, except for an excess of high-stellar mass, high star forming galaxies related to the large program studying luminous-infrared galaxies which are preferentially found among the highest star forming and higher stellar mass galaxies in the nearby universe\cite{U:2012:9}.  As galaxy mergers are thought to correlate with increased star formation, the lack of nuclear mergers in these inactive galaxies is very surprising and strengthens our findings of an excess of (nuclear) mergers among the luminous obscured AGN population, compared to a control sample.
    
Finally, when comparing merger fractions between samples we use the binomial proportion confidence intervals (CI) are typically used to compare the fraction of different samples.  The normal approximation interval is the simplest formula, however for situations with a fraction very close to zero or small numbers this formula is unreliable\cite{DasGupta:2001:101} and may significantly underestimate the uncertainties.  We therefore use Jeffrey's confidence interval to provide more reliable error estimates and Fisher's exact test to calculate the $p$-value significance of the difference in the two sample proportions.  


\noindent{\bf Simulations of galaxies at high redshift.} We simulate the systematics of studying AGN at the peak of black hole growth at higher redshift ($z\approx1$) by artificially redshifting our imaging data to mimic the data quality of \H\ at this redshift following Ref.~\citen{Hung:2014:63}.  Because of the low redshift of our samples ($\overline{z}=0.04$), the physical resolution of our ground-based Pan-STARRS images is equal or superior to \H data for a $z{\approx}1$ sample.  The images also have complete wavelength coverage in the $g$, $r$, $i$, $z$, and $y$ filters ($\lambda_{\rm eff}$ =4776, 6130, 7485, 8658, and 9603\AA, respectively), and thus we can properly consider the rest-frame wavelengths and spectral energy distributions (SEDs) of the artificially redshifted datasets.  We assume the redshifted galaxies are located at $z{=}1$, and observed with WFC3 F160W, achieving the same depth as the CANDELS GOODS-S surveys\cite{Grogin:2011:35}.  $\mathtt{FERENGI}$ then determines the best-fit rest-frame SED templates using the $\mathtt{kcorrect}$ routine, and calculates the expected flux in the WFC3 F160W band. Finally, the output spatial flux distribution is convolved with the PSF of the WFC3 F160W band, and a noise frame is added using a blank region extracted from the CANDELS GOODS-S surveys.

Our simulated observations show that \hst is not able to resolve all but one of these late stage mergers with tight double nuclei.  This is not surprising, given the stark difference in physical scales between $z{\approx}0.04$ and $z\approx1$ ( $D_{A\,(z{=}0.04)}/D_{A\,(z{=}1)}{\approx} 10$) and the detection based on the visibility of tight double nuclei. While our asymmetric/disturbed structures at the outskirts of the galaxies may still be visible at $z{\approx}1$ in a couple of cases with a significantly increased level of brightness due to the increase in star formation activity at high redshift, the interpretation of these structures can be ambiguous given that the morphology of high-redshift star-forming galaxies are often intrinsically less regular.

\noindent{\bf Simulations of merging galaxies.} We use high-resolution galaxy merger simulations performed with {\footnotesize GADGET}, a smoothed-particle hydrodynamics (SPH) and N-body code that conserves energy and entropy and uses sub-resolution physical models for radiative heating and cooling, star formation, supernova feedback, metal enrichment, and a multi-phase interstellar medium\cite{Springel:2003:289,Springel:2005:1105}. 
BHs are modelled as gravitational ``sink" particles that accrete gas via an Eddington-limited, Bondi-Hoyle like prescription. Thermal AGN feedback is included by coupling 5\% of the accretion luminosity ($L_{\rm bol} {=} \epsilon_{\rm rad} \dot{M} c^2$) to the surrounding gas as thermal energy, with a variable radiative efficiency $\epsilon_{\rm rad}$ at low accretion rates\cite{Narayan:2008:733}.


Our simulation suite includes 7 major-merger simulations with galaxy mass ratios of 0.5 or 1.  The galaxies each consist of a dark matter halo, a disk of gas and stars (with initial gas fractions of 0.1-0.3), a stellar bulge-to-total ratio of 0 or 0.2, and a central BH with initial mass scaled to the stellar bulge\cite{Kormendy:2013:511}. 
The fiducial baryonic gravitational softening length and mass resolution are $\epsilon_{\rm grav} =48$ pc and $m_{\rm b} = 2.8\times 10^5$ \Msun, respectively. We have also run two simulations at ten times higher mass resolution, to ensure that our results are not resolution-dependent.  We stress that these details are not crucial for the purpose of the present work, in which the simulations are used to asses the relative time-scales during which merging galaxy nuclei and BHs can be seen at various separations.


We also conduct radiative transfer simulations in post-processing with the 3-D, polychromatic, dust radiative transfer code \sunrise \cite{Jonsson:2006:2,Jonsson:2010:17}. This publicly-available code has been used extensively with \gadget\ to model a wide range of isolated and merging galaxy populations\cite{Narayanan:2010:1701, Snyder:2013:168, Blecha:2013:1341}. Stellar emission is calculated from age- and metallicity-dependent {\footnotesize STARBURST99} SEDs for each stellar particle\cite{Leitherer:1999:3} and emission from H~{\sc ii} regions (including dusty photo-dissociation regions) around young stars is calculated using the {\footnotesize MAPPINGSIII} models\cite{Groves:2008:438}. An AGN SED is implemented based on the BH accretion rate; our fiducial model is based on empirically-derived, luminosity-dependent templates\cite{Hopkins:2007:731}.

After the dust distribution is calculated in \sunrise\ from the gas-phase metal density distribution, \sunrise\ performs Monte Carlo radiative transfer through the dust grid, computing energy absorption (including dust self-absorption) and thermal re-emission to produce the emergent, spatially-resolved UV-to-IR SEDs. For each merger simulation, we run \sunrise\ on snapshots at 10 Myr intervals during the merger phase (separations $<$ 10-30 kpc) and post-merger phases, and at 100 Myr intervals during the early-merger phase, for seven isotropically-distributed viewing angles and the result is converted to a merger fraction as would be seen if observed from a single direction. 

\noindent{\bf Code availability.} The custom NIRC2 reduction software is available at https://github.com/jluastro/JLU-python-code/tree/master/jlu.

\noindent{\bf Data availability.}  The reduced imaging datasets from the \hst are available from the Hubble Legacy Archive. The raw imaging datasets from the near-infrared adaptive optics programmes are available from the Keck Observatory Archive. Other reduced datasets generated or analysed in this study are available from the corresponding author on reasonable request.\\



\begin{addendum}
\item[Acknowledgements] 
This work is dedicated to the memory of our friend and
collaborator N. Gehrels. M.J.K. acknowledges support from the Swiss National
Science Foundation (SNSF) through the Ambizione fellowship grant PZ00P2
154799/1 and from NASA through ADAP award NNH16CT03C. K.S., L.F.S. and
A.W. acknowledge support from SNSF grants PP00P2 138979 and PP00P2
166159.   L.B. acknowledges support from NSF award number 1715413. We acknowledge the work of the Swift/BAT team to make this study
possible. This paper is part of the Swift/BAT AGN Spectroscopic Survey (BASS, https:/www.bass-survey.com).

\item[Reviwer information] 
Nature thanks D. Kocevski and the other anonymous reviewer(s) for their contribution to the peer review of this work.

\item[Author Contributions]
M.J.K. drafted the manuscript, performed the observations
and carried out much of the analysis. L.B. performed and interpreted the
hydrodynamic simulations. P.B. carried out much of the initial data reduction.
C.-L.H. ran the artificial redshifting code. J.R.L. provided the initial data reduction
code and helped with the analysis. K.S. aided in the scientific interpretations and the reduction of the raw data. E.T. assisted in the initial observing runs. R.M.,
S.V. and D.B.S. aided in the initial proposal and scientific interpretations. B.T.,
L.F.S., A.W. and C.R. assisted in the scientific interpretations.

\item[Author Information] 
The authors declare no competing financial interests. Readers are welcome to comment on the online version of the paper. Correspondence and requests for materials should be addressed to M.K.~(mike.koss@eurekasci.com).

\item[Additional Information] 
Extended data is available for this paper at https://doi.org/10.1038/s41586-018-0652-7. Supplementary information is available for this paper at https://doi.org/10.1038/s41586-018-0652-7.
 
\end{addendum}

\clearpage

\renewcommand{\thefigure}{Extended \arabic{figure}}
\renewcommand{\thetable}{Extended \arabic{table}}
\setcounter{figure}{0}

\begin{figure*} 
\centering
\includegraphics[height=23cm]{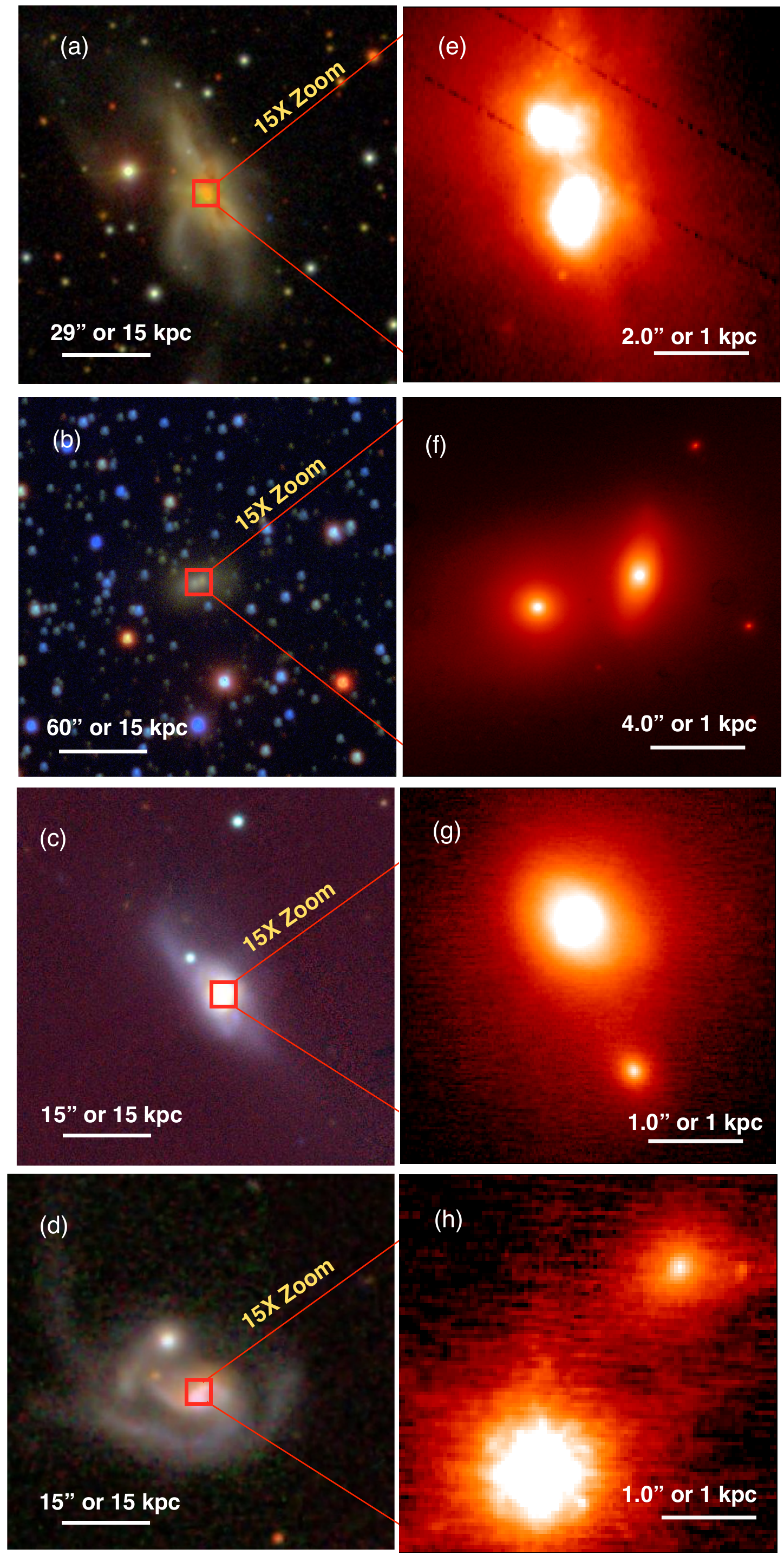}
\caption*{Extended Data Figure 1: \textbf{Other close mergers}.  \textbf{a-d,} Tricolor optical images in $gri$ from from SDSS or Kitt Peak imaging with about $1\arcsec$ angular resolution.  The galaxies shown are NGC 6240 (\textbf{a}), 2MASX J00253292+6821442 (\textbf{b}), ESO 509-G027 (\textbf{c}) and Mrk 975 (\textbf{d}) from the AGN sample.   The images are 60 kpc by 60 kpc in size.  Red squares indicate the size of the zoomed in AO image on the right.  \textbf{e-h,}  High spatial resolution images of nuclear mergers shown in \textbf{a-d,} 4 kpc by 4 kpc in size.  
}
\label{mergersadd2}
\end{figure*}

\begin{figure*} 
\centering
\includegraphics[height=22cm]{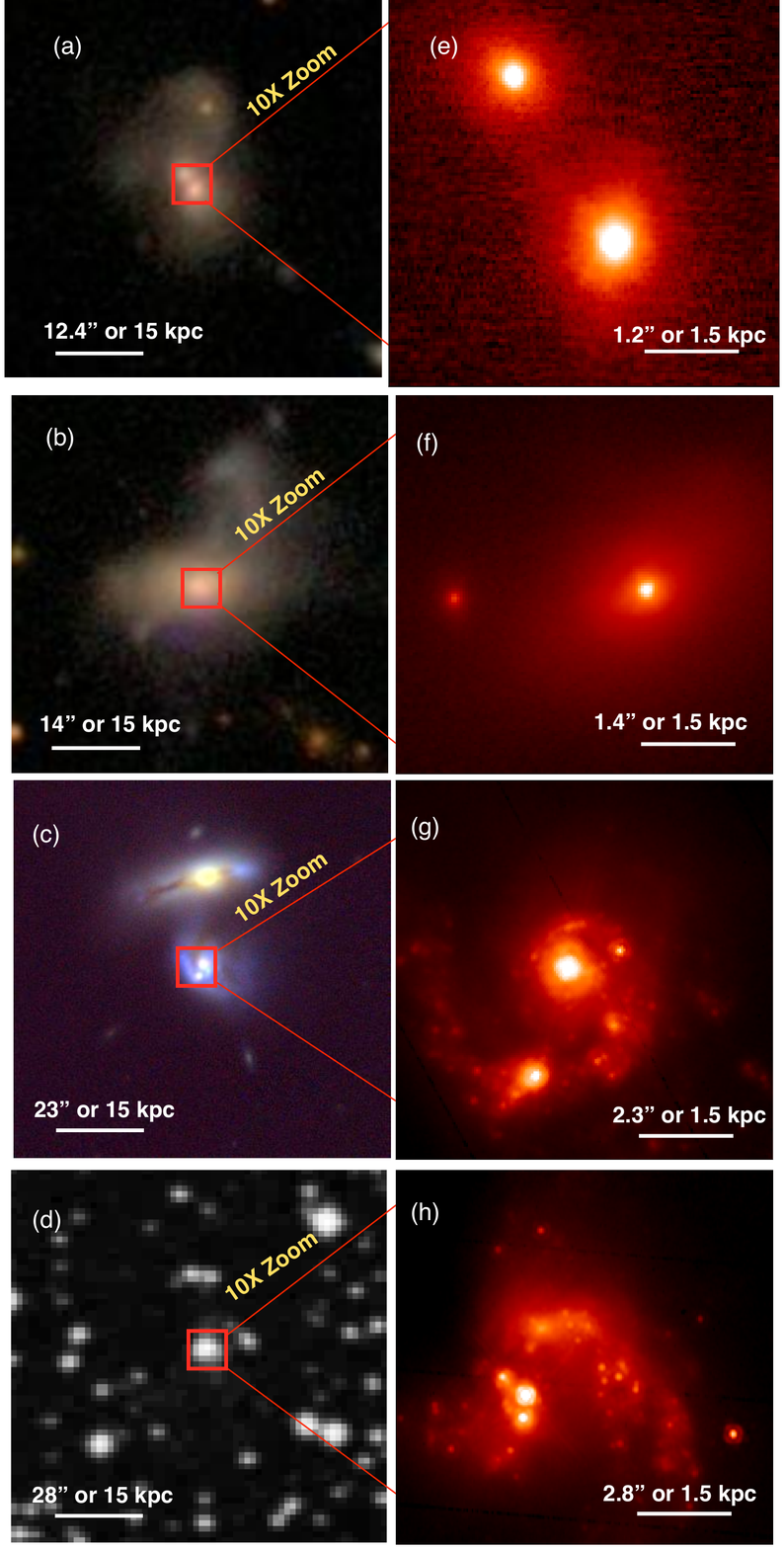}
\caption*{Extended Data Figure 2: \textbf{Other Close Mergers Continued}.  \textbf{a-c,} Tricolor optical images in $gri$ from from SDSS or Kitt Peak imaging with about 1\arcsec\ angular resolution.  The galaxies shown are 2MASX J16311554+2352577 (\textbf{a}) and 2MASX J08434495+3549421 (\textbf{b}) from the AGN sample and 2MASX J08370182-4954302 (\textbf{c}) from the inactive-galaxy sample. \textbf{d}, Lower-quality red Digitized Sky Survey image of UGC02369 NED01, for which no higher-quality imaging exists. The images in \textbf{a-d} are  60 kpc by 60 kpc in size.  Red squares indicate the size of the zoomed in AO image on the right.  \textbf{e-h,}  High-spatial-resolution near-infrared images of the nuclear mergers shown in \textbf{a-d}, 4 kpc by 4 kpc in size.}
\label{mergersadd2}
\end{figure*}

\begin{figure*} 

\centering
\includegraphics[height=22cm]{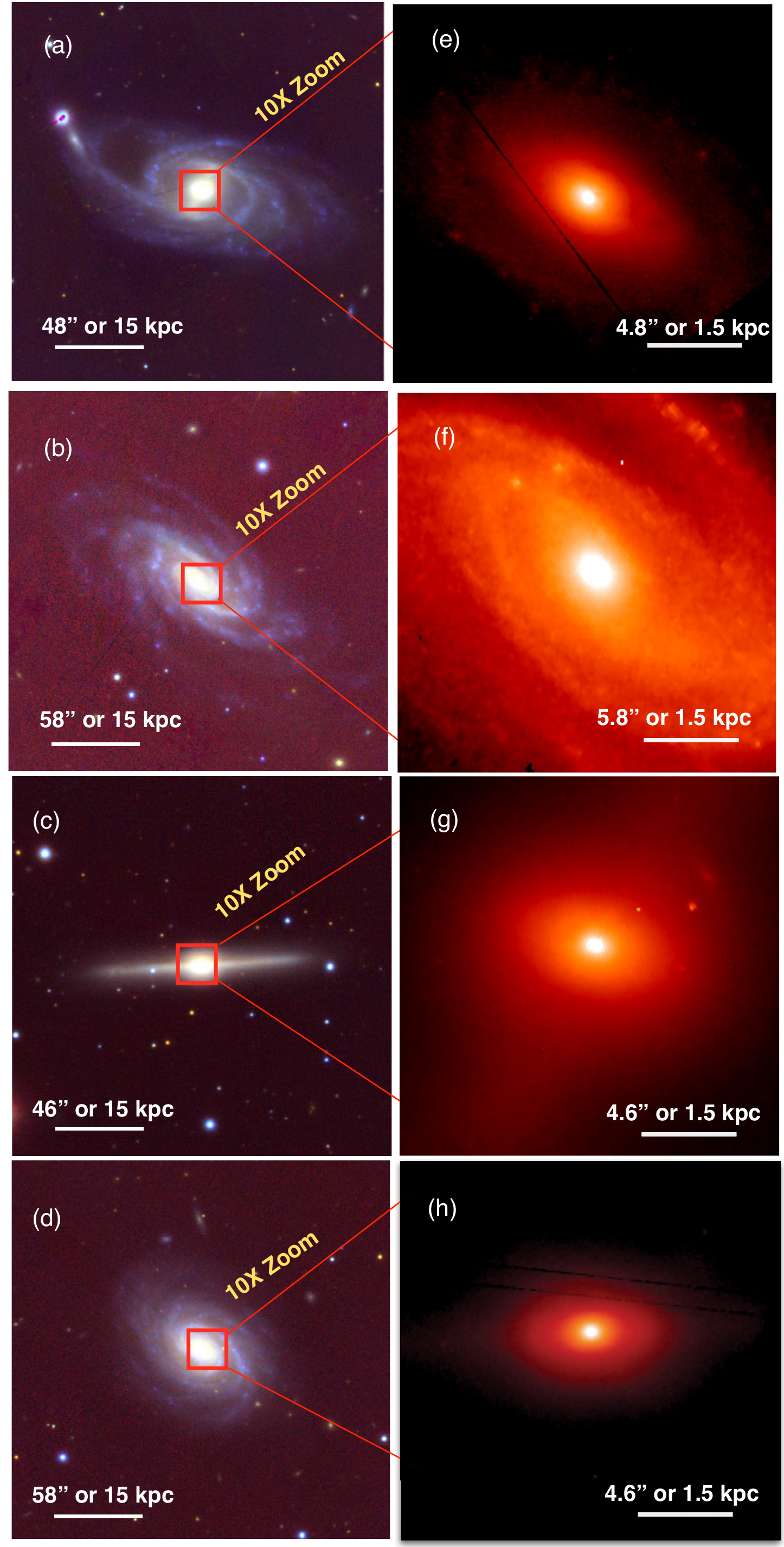}

\caption*{Extended Data Figure 3:  \textbf{Inactive-galaxy control sample}. \textbf{a-d,} Tricolor optical images in the $gri$ band from Pan-Starrs imaging with about 1\arcsec angular resolution. The images show inactive galaxies in the control sample that were matched in stellar mass and SFR to the AGN: NGC 214 (\textbf{a}), NGC 151 (\textbf{b}), NGC 2998 (\textbf{c}) and NGC 6504 (\textbf{d}). The images are 60 kpc by 60 kpc in size.  Red squares indicate the size of the zoomed in AO image on the right.  \textbf{e-h,}  High-spatial-resolution near-infrared images of the nuclear mergers shown in \textbf{a-d}, 4 kpc by 4 kpc in size.  Some white lines are present in NICMOS and Pan-STARRS imaging owing to bad pixels with very low or zero response or with very high or erratic dark current.}
\label{control}
\end{figure*}  

\begin{figure} 

\centering
\includegraphics[width=8.5cm]{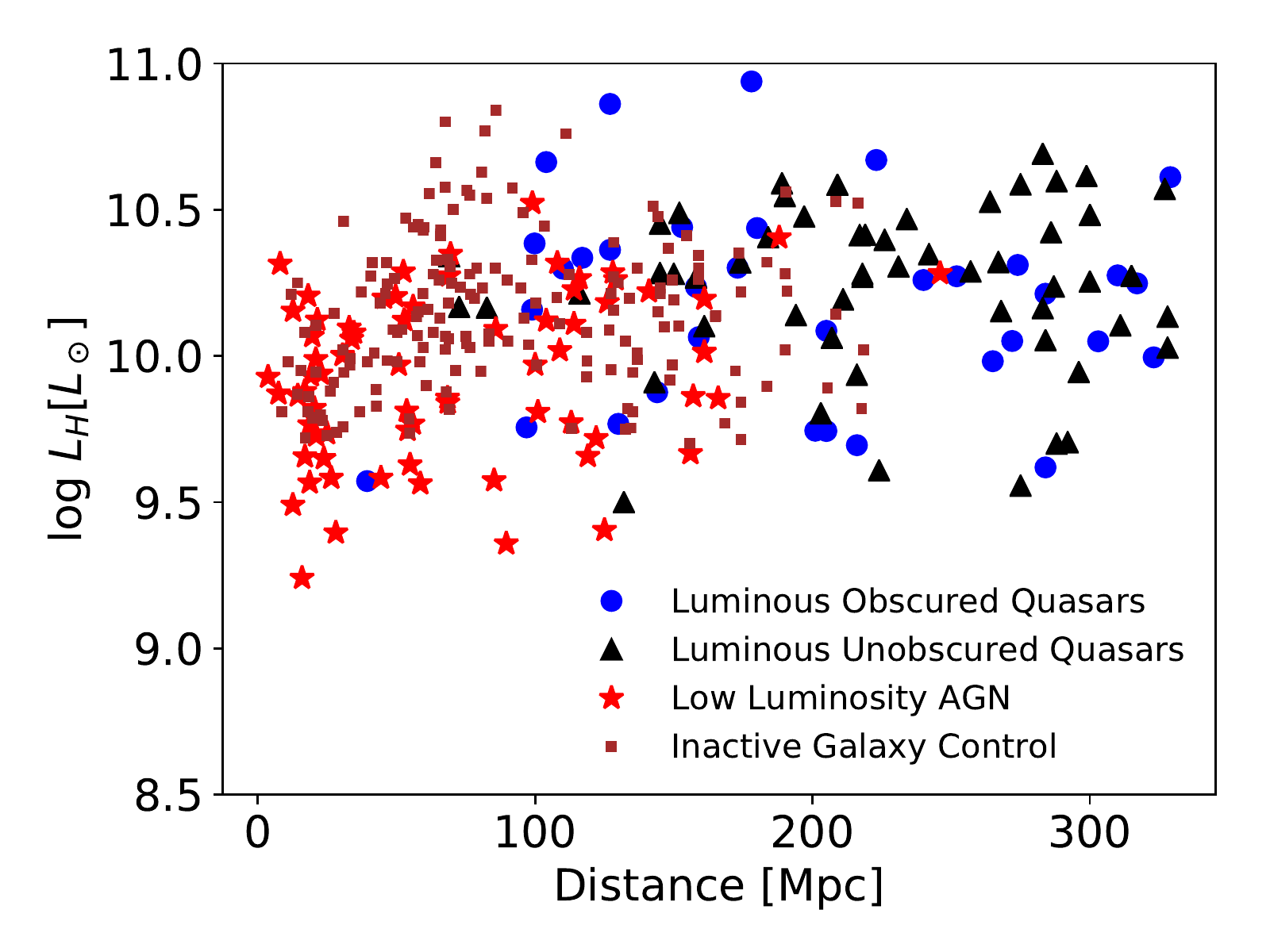}
\includegraphics[width=8.5cm]{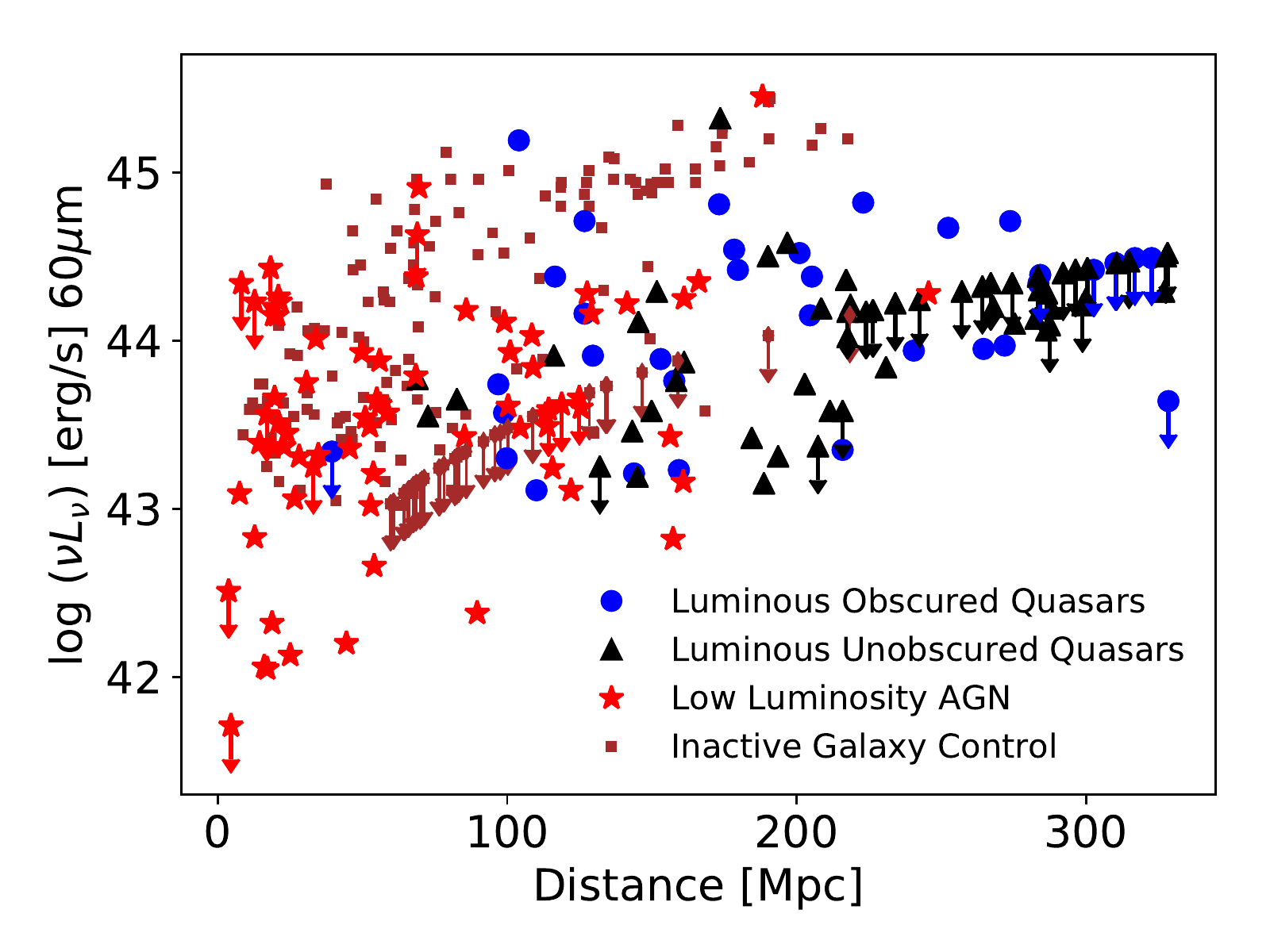}
\includegraphics[width=8.5cm]{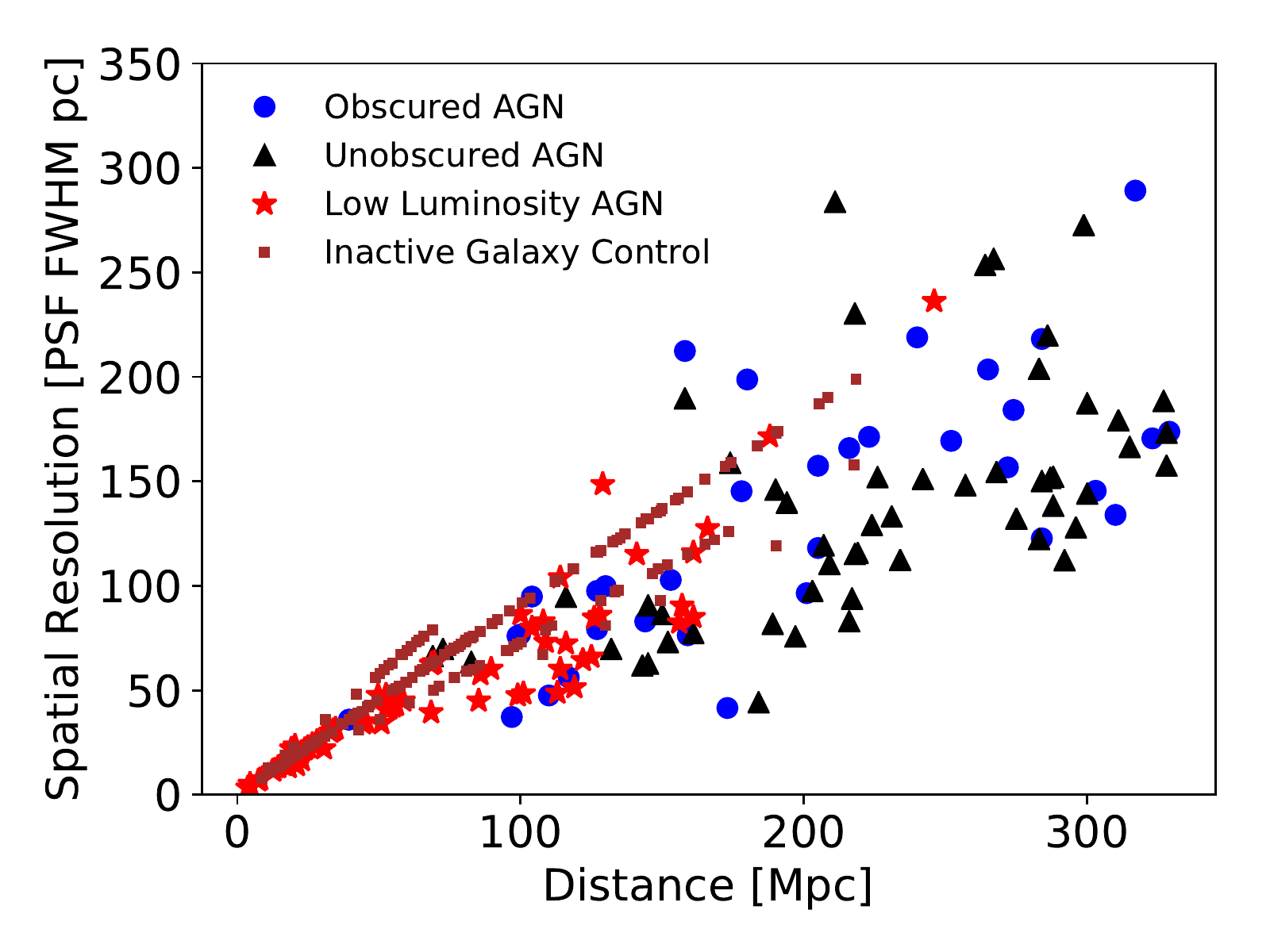}
\caption*{Extended Data Figure 4: \textbf{Stellar mass, star formation rate and resolution of AGN and inactive galaxies.} {\textbf a,} $H$-band luminosity of the different AGN and inactive galaxies.  Inactive galaxies with considerably lower stellarmasses than the AGN samples were excluded  ($\log$ $L_H [L_\odot]<9.7$).  {\textbf b,}  60-\micron\ luminosity of the different AGN and inactive galaxies. Inactive galaxies with lower SFR were also excluded from the comparison ($\log$ ($\nu L_{\nu}$) [erg/s] 60$\mu$m$<43$). For observations in which a galaxy was not detected, we show a 3$\sigma$ upper limit of the SFR, indicated by a downward arrow.     {\textbf c,} Comparison of the maximum spatial resolution (in parsecs) of the different observations. The inactive-galaxy sample typically has higher physical spatial resolutions than the AGN samples. Many galaxies observed fall along a line because of the constant physical resolution of the \hstsh. }
\label{control_compstellar}
\end{figure}

\begin{figure} 

\centering
\includegraphics[width=9cm]{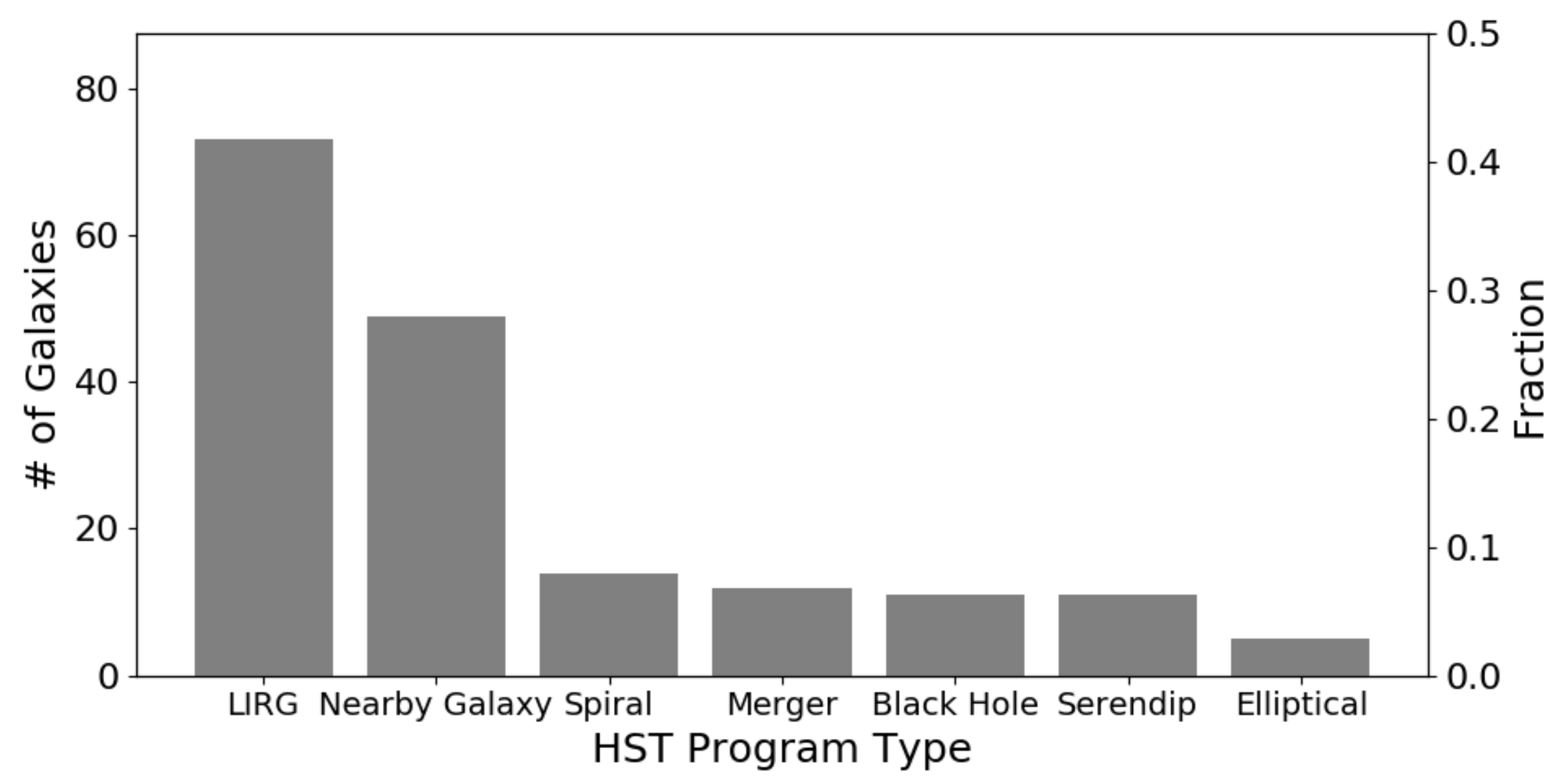}
\caption*{Extended Data Figure 5: \textbf{Summary of programme types included in the \hst control sample.}  The majority of archival control sample observations are of high-SFR luminous infrared galaxies (‘LIRG’) or from studies of volume-limited samples of nearby galaxies (‘Nearby Galaxy’). The remaining samples originate from observations of spiral galaxies (‘Spiral’), galaxies in the merger sequence or late-stage mergers (‘Merger’), galaxies with large or small black holes (‘Black Hole’) and elliptical galaxies (‘Elliptical’). Finally, some nearby galaxies were observed serendipitously in observations of other sources or survey fields (‘Serendip’).
}
\label{mergersadd2}
\end{figure} 

\begin{figure} 

\centering
\includegraphics[width=9cm]{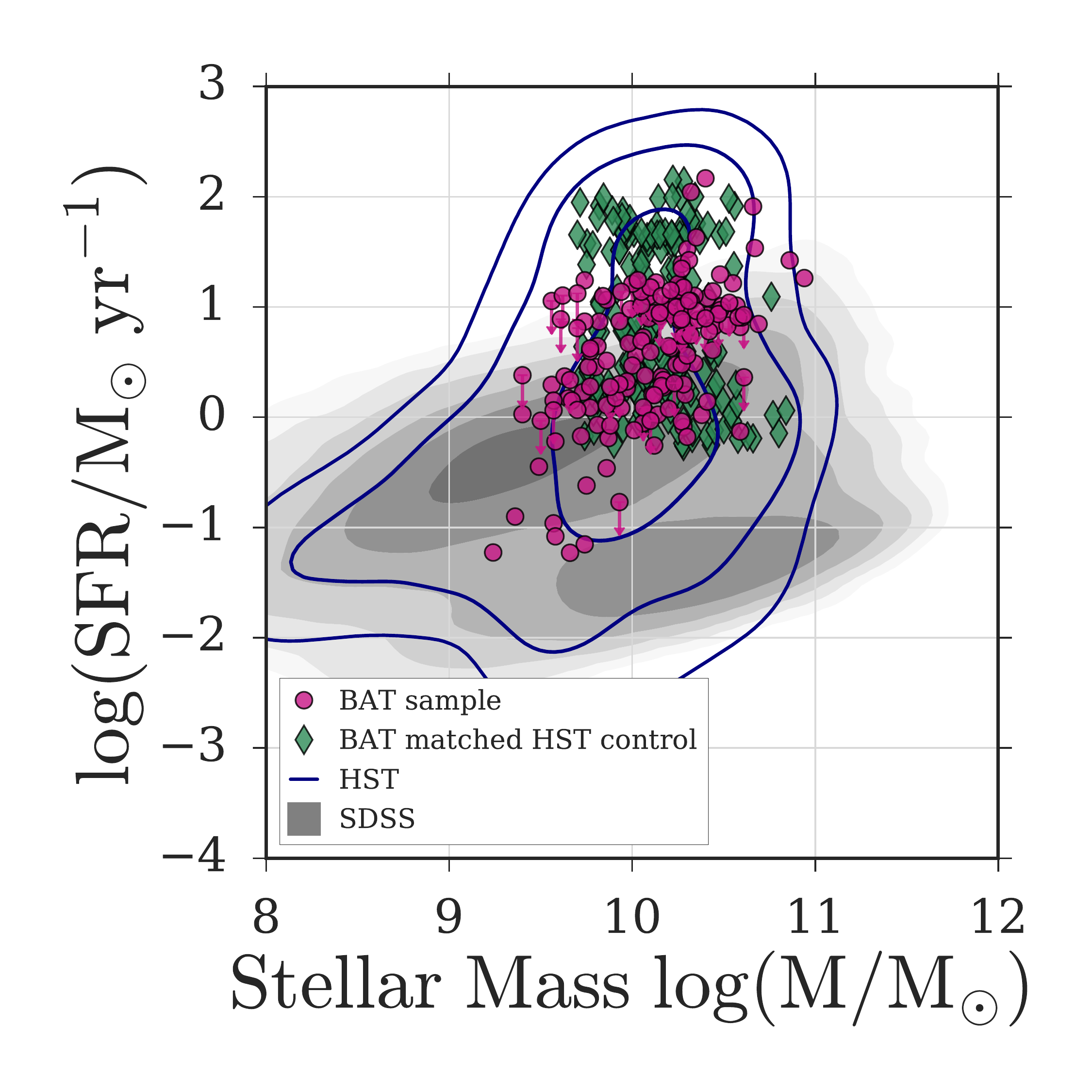}
\caption*{Extended Data Figure 6: \textbf{SFR and stellar mass.} Measurements of SFR and stellar mass for the BAT AGN sample (purple circles) and the \hstsh-matched archival control sample of inactive galaxies (green diamonds). The full distribution of inactive galaxies from the Sloan Digital Sky Survey (SDSS) is shown with grey shading and the full distribution of the \hst archive with blue contours. The HST archival sample has an excess of high-stellar-mass, high-SFR inactive galaxies because of the large number of observations of luminous infrared galaxies.
}
\label{mergersadd2}
\end{figure}  

\begin{figure}
\centering
\includegraphics[width=5.3cm]{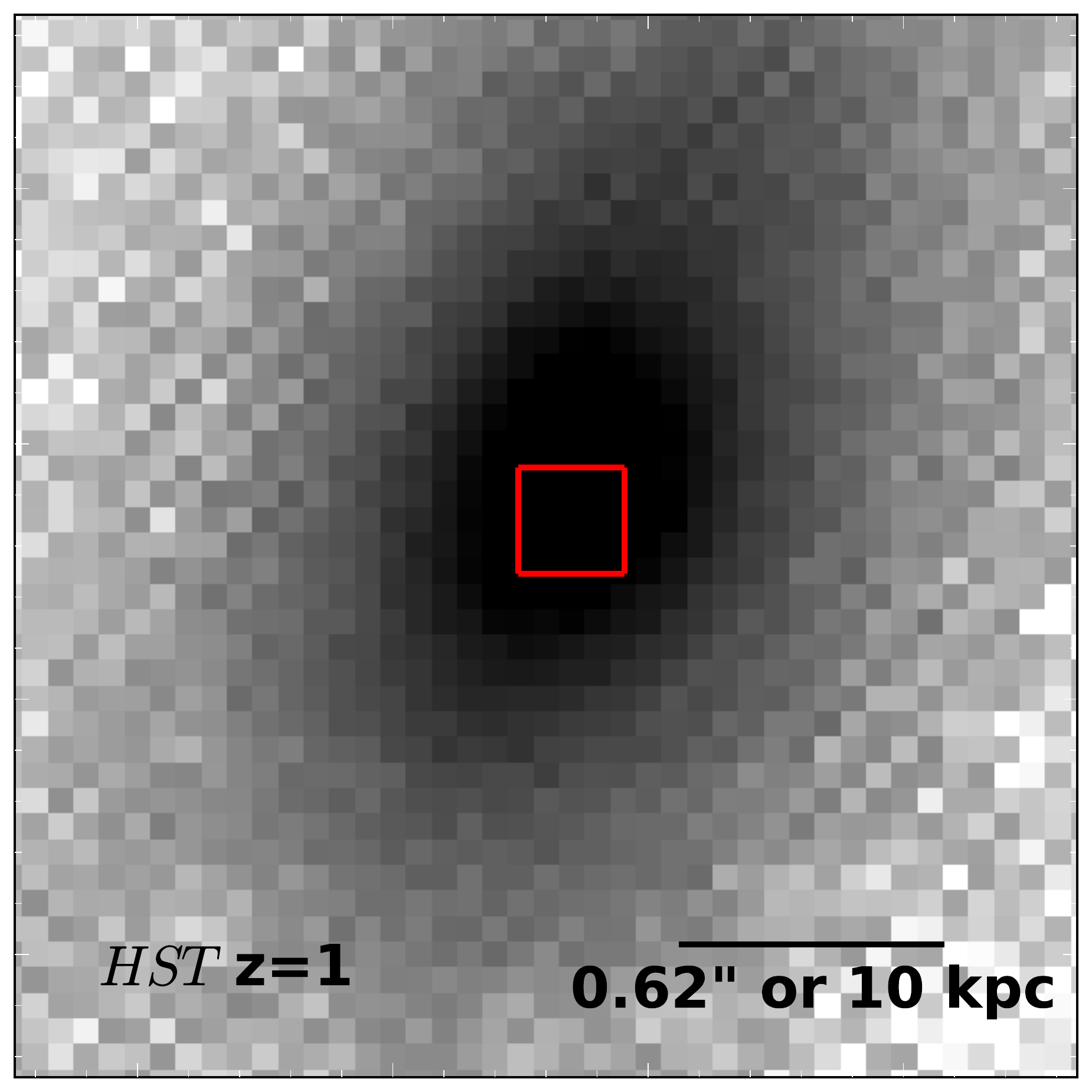}
\includegraphics[width=5.3cm]{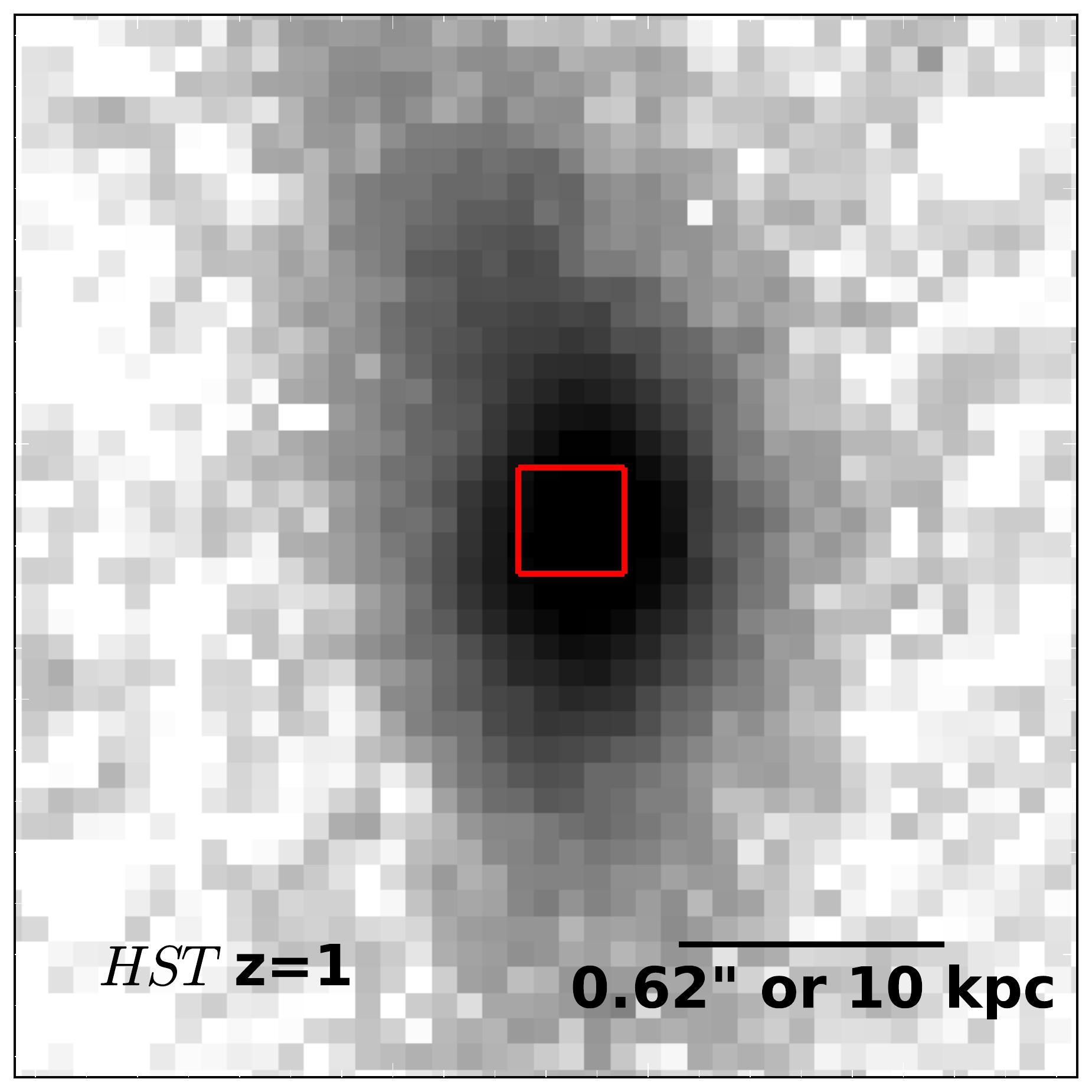}
\includegraphics[width=5.3cm]{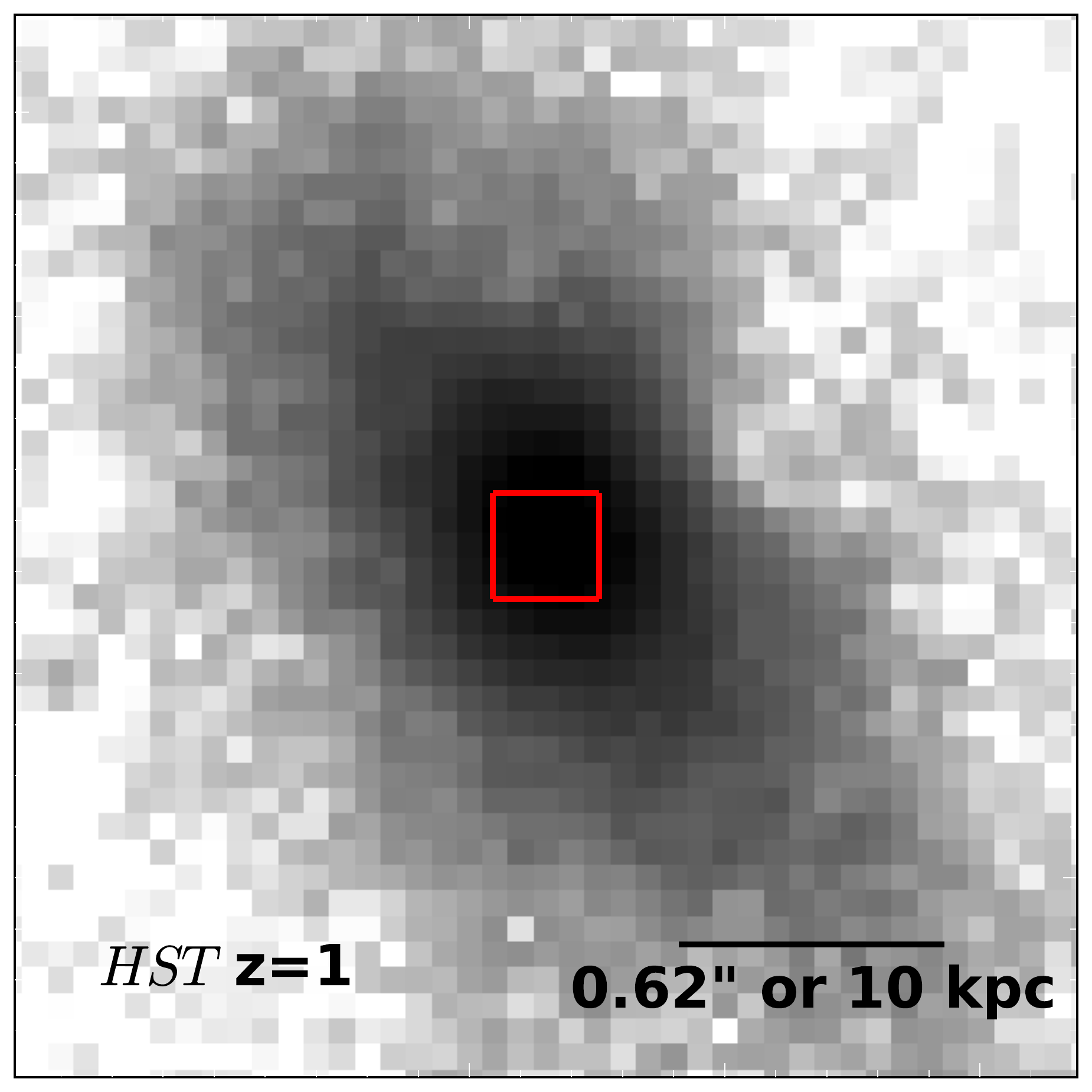}
\caption*{Extended Data Figure 7: \textbf{Simulated \hst images of nuclear mergers at high redshift.}  Simulated images of three nuclear mergers (2MASXJ 01392400+2924067, CGCG 341-006, MCG+02-21-013) observed at $z=1$ with $HST$ F160W as part of CANDLES survey (60 mas pixel$^{-1}$) using optical imaging and {\sc Ferengi} software.  The \hst would be unable to detect these final stage mergers. All simulated images are displayed in the arcsinh scale in coupled-channel-device counts, as if observed in the \hst F160W filter as part of the CANDELS survey.}
\label{hidden}
\end{figure}

\clearpage
\onecolumn

\begin{center}
\begin{footnotesize}
\begin{longtable}{| l | l | l | l | l | l | l |}
\caption{\textbf{Extended Data Table 1:Galaxies with companions within 10 kpc}}
\label{allobs}\\

\hline
\multicolumn{1}{|c|}{\textbf{Galaxy}} & \multicolumn{1}{c|}{\textbf{Class}} & \multicolumn{1}{c|}{\textbf{d}} & \multicolumn{1}{c|}{\textbf{d}} & \multicolumn{1}{c|}{\textbf{Stellarity}} & \multicolumn{1}{c|}{\textbf{Stellar Contam}} & \multicolumn{1}{c|}{\textbf{Diff}}\\ 
\multicolumn{1}{|c|}{\textbf{}} & \multicolumn{1}{c|}{\textbf{}} & \multicolumn{1}{c|}{\textbf{[\arcsec]}} & \multicolumn{1}{c|}{\textbf{[kpc]}} & \multicolumn{1}{c|}{\textbf{[\arcsec]}} & \multicolumn{1}{c|}{\textbf{\%}} & \multicolumn{1}{c|}{\textbf{[mag]}}\\ \hline 
\endfirsthead

\hline
\endlastfoot
MCG+02-21-013&LumObs&0.4&0.3&0&--&1.4\\
NGC 6240&LumObs&1.8&0.9&0&--&0.8\\
2MASXJ08370182-4954302&Inactive&2.1&1.1&--&--&1.0\\
2MASX J00253292+6821442&LowLum&4.5&1.1&0&--&0.3\\
CGCG341-006&LumObs&1.3&1.3&0&--&0.6\\
UGC02369NED01&Inactive&3&1.9&--&--&0.5\\
2MASXJ01392400+2924067&LumObs&1.2&1.9&0&--&1.1\\
Mrk975&LumUnob&2.5&2.5&0&3.2&0.9\\
2MASXJ16311554+2352577&LumObs&2.3&2.8&0&--&0.8\\
2MASXJ08434495+3549421&LumObs&2.7&2.9&0&--&2.4\\
ESO099-G004&Inactive&5.2&3.1&--&--&1.6\\
MCG+12-02-001&Inactive&9.6&3.1&--&--&0.4\\
NGC985&LumUnob&3.7&3.3&0&--&2.1\\
IRAS23436+5257&Inactive&5.3&3.6&--&--&0.4\\
MCG-02-33-098&Inactive&11.4&3.6&--&--&0.1\\
Mrk739E&LowLum&6.1&3.8&0&--&0.3\\
NGC6090NED02&Inactive&6.6&3.9&--&--&1.3\\
NGC3588NED01&LowLum&7.6&4.1&0&--&1.1\\
Mrk463&LumObs&4.8&5.0&0&--&1.4\\
2MASXJ06094582-2140234&Inactive&8.1&6.1&--&--&1.2\\
Mrk423&LowLum&9.1&6.1&0&--&1.7\\
2MASXJ05442257+5907361&LumObs&4.8&6.6&0&--&2.2\\
IRAS21101+5810&Inactive&9.4&7.3&--&--&0.8\\
IIZw096NED02&Inactive&10.2&7.4&--&--&0.5\\
IRASF03359+1523&Inactive&10.8&7.7&--&--&0.2\\
NGC7212NED02&LowLum&14.3&7.9&0&--&0.9\\
Was49b&LumObs&6.7&8.7&0&--&0.7\\
NGC2672&Inactive&28.2&9.2&0.11&--&1.2\\
UGC04881&Inactive&11.3&9.0&--&--&0.3\\
2MASXJ17085915+2153082&LumUnob&6.7&9.8&0&--&1.3\\
\hline
\caption{The table lists the sources found to have counterparts within 10 kpc. Obscured and unobscured AGN are separated using the presence of broad H$\beta$ lines in their optical spectra, and the separation between low- and high-luminosity AGN (below or above $L_{bol} = 2 \times 10^{44}$ \ergps, respectively) is based on their X-ray emission (‘LowLum’, low-luminosity AGN; ‘LumUnob’, luminous unobscured AGN;
‘LumObs’, luminous obscured AGN). The separation d between the two galaxy nuclei is given in arcseconds and kiloparsecs. The stellar contamination (‘Stellar Contam’) indicates the likelihood of a stellar source of this brightness occurring randomly in the same search area. Finally, the measured stellarity index from a neural net (‘Stell.’) and the difference (in mag) between the primary and secondary galaxies in the merger (‘Diff mag’) are also listed.}
\end{longtable}

\end{footnotesize}
\end{center}

\noindent \textbf{Supplementary Data}
The first table, All High Resolution Observations, contains a list of all the galaxies in the study and the details of their high resolution observations, $H$-band emission as a proxy for star formation and 60 \micron\ emission as a proxy for their star formation.  The second table, \hst archival programs, contains a list of all the \hst archival programs used in the study. 


\noindent {\bf References}
\begin{thebibliography}{10}
\expandafter\ifx\csname url\endcsname\relax
  \def\url#1{\texttt{#1}}\fi
\expandafter\ifx\csname urlprefix\endcsname\relax\def\urlprefix{URL }\fi
\providecommand{\bibinfo}[2]{#2}
\providecommand{\eprint}[2][]{\url{#2}}

\bibitem{DiMatteo:2005:604}
\bibinfo{author}{Di~Matteo, T.}, \bibinfo{author}{Springel, V.} \&
  \bibinfo{author}{Hernquist, L.}
\newblock {Energy input from quasars regulates the growth and activity of black
  holes and their host galaxies}.
\newblock \emph{\bibinfo{journal}{Nature}} \textbf{\bibinfo{volume}{433}},
  \bibinfo{pages}{604} (\bibinfo{year}{2005}).

\bibitem{Goulding:2018:S37}
\bibinfo{author}{Goulding, A.~D.} \emph{et~al.}
\newblock {Galaxy interactions trigger rapid black hole growth: An
  unprecedented view from the Hyper Suprime-Cam survey}.
\newblock \emph{\bibinfo{journal}{Publ. Astron. Soc. JPN}}
  \textbf{\bibinfo{volume}{70}}, \bibinfo{pages}{S37} (\bibinfo{year}{2018}).

\bibitem{Donley:2018:63}
\bibinfo{author}{Donley, J.~L.} \emph{et~al.}
\newblock {Evidence for Merger-driven Growth in Luminous, High-z, Obscured AGNs
  in the CANDELS/COSMOS Field}.
\newblock \emph{\bibinfo{journal}{Astrophys. J.}}
  \textbf{\bibinfo{volume}{853}}, \bibinfo{pages}{63} (\bibinfo{year}{2018}).

\bibitem{Villforth:2017:812}
\bibinfo{author}{Villforth, C.} \emph{et~al.}
\newblock {Host galaxies of luminous z {\~{ }} 0.6 quasars: major mergers are
  not prevalent at the highest AGN luminosities}.
\newblock \emph{\bibinfo{journal}{Mon. Not. R. Astron. Soc.}}
  \textbf{\bibinfo{volume}{466}}, \bibinfo{pages}{812--830}
  (\bibinfo{year}{2017}).

\bibitem{Chang:2017:19}
\bibinfo{author}{Chang, Y.-Y.} \emph{et~al.}
\newblock {Infrared Selection of Obscured Active Galactic Nuclei in the COSMOS
  Field}.
\newblock \emph{\bibinfo{journal}{Astrophys. J. Suppl. Ser.}}
  \textbf{\bibinfo{volume}{233}}, \bibinfo{pages}{19} (\bibinfo{year}{2017}).

\bibitem{Kocevski:2015:104}
\bibinfo{author}{Kocevski, D.~D.} \emph{et~al.}
\newblock {Are Compton-thick AGNs the Missing Link between Mergers and Black
  Hole Growth?}
\newblock \emph{\bibinfo{journal}{Astrophys. J.}}
  \textbf{\bibinfo{volume}{814}}, \bibinfo{pages}{104} (\bibinfo{year}{2015}).

\bibitem{Koss:2016:85}
\bibinfo{author}{Koss, M.~J.} \emph{et~al.}
\newblock {A New Population of Compton-thick AGNs Identified Using the Spectral
  Curvature above 10 keV}.
\newblock \emph{\bibinfo{journal}{Astrophys. J.}}
  \textbf{\bibinfo{volume}{825}}, \bibinfo{pages}{85} (\bibinfo{year}{2016}).

\bibitem{Ricci:2017:stx173}
\bibinfo{author}{Ricci, C.} \emph{et~al.}
\newblock {Growing supermassive black holes in the late stages of galaxy
  mergers are heavily obscured}.
\newblock \emph{\bibinfo{journal}{Mon. Not. R. Astron. Soc.}}
  \bibinfo{pages}{stx173} (\bibinfo{year}{2017}).

\bibitem{Hopkins:2005:L71}
\bibinfo{author}{Hopkins, P.~F.} \emph{et~al.}
\newblock {A Physical Model for the Origin of Quasar Lifetimes}.
\newblock \emph{\bibinfo{journal}{Astrophys. J.}}
  \textbf{\bibinfo{volume}{625}}, \bibinfo{pages}{L71--L74}
  (\bibinfo{year}{2005}).

\bibitem{Baumgartner:2013:19}
\bibinfo{author}{Baumgartner, W.~H.} \emph{et~al.}
\newblock {The 70 Month Swift-BAT All-sky Hard X-Ray Survey}.
\newblock \emph{\bibinfo{journal}{Astrophys. J. Suppl. Ser.}}
  \textbf{\bibinfo{volume}{207}}, \bibinfo{pages}{19} (\bibinfo{year}{2013}).

\bibitem{Koss:2017:74}
\bibinfo{author}{Koss, M.} \emph{et~al.}
\newblock {BAT AGN Spectroscopic Survey. I. Spectral Measurements, Derived
  Quantities, and AGN Demographics}.
\newblock \emph{\bibinfo{journal}{Astrophys. J.}}
  \textbf{\bibinfo{volume}{850}}, \bibinfo{pages}{74} (\bibinfo{year}{2017}).

\bibitem{Ricci:2017:17}
\bibinfo{author}{Ricci, C.} \emph{et~al.}
\newblock {BAT AGN Spectroscopic Survey. V. X-Ray Properties of the Swift/BAT
  70-month AGN Catalog}.
\newblock \emph{\bibinfo{journal}{Astrophys. J. Suppl. Ser.}}
  \textbf{\bibinfo{volume}{233}}, \bibinfo{pages}{17} (\bibinfo{year}{2017}).

\bibitem{Ohyama:2015:162}
\bibinfo{author}{Ohyama, Y.}, \bibinfo{author}{Terashima, Y.} \&
  \bibinfo{author}{Sakamoto, K.}
\newblock {Infrared and X-Ray Evidence of an AGN in the NGC 3256 Southern
  Nucleus}.
\newblock \emph{\bibinfo{journal}{Astrophys. J.}}
  \textbf{\bibinfo{volume}{805}}, \bibinfo{pages}{162} (\bibinfo{year}{2015}).

\bibitem{Barrows:2017:129}
\bibinfo{author}{Barrows, R.~S.}, \bibinfo{author}{Comerford, J.~M.},
  \bibinfo{author}{Greene, J.~E.} \& \bibinfo{author}{Pooley, D.}
\newblock {Spatially Offset Active Galactic Nuclei. II. Triggering in Galaxy
  Mergers}.
\newblock \emph{\bibinfo{journal}{Astrophys. J.}}
  \textbf{\bibinfo{volume}{838}}, \bibinfo{pages}{129} (\bibinfo{year}{2017}).

\bibitem{Fu:2011:103}
\bibinfo{author}{Fu, H.}, \bibinfo{author}{Myers, A.~D.},
  \bibinfo{author}{Djorgovski, S.~G.} \& \bibinfo{author}{Yan, L.}
\newblock {Mergers in Double-peaked [O III] Active Galactic Nuclei}.
\newblock \emph{\bibinfo{journal}{Astrophys. J.}}
  \textbf{\bibinfo{volume}{733}}, \bibinfo{pages}{103} (\bibinfo{year}{2011}).

\bibitem{Haan:2011:100}
\bibinfo{author}{Haan, S.} \emph{et~al.}
\newblock {The Nuclear Structure in Nearby Luminous Infrared Galaxies: Hubble
  Space Telescope NICMOS Imaging of the GOALS Sample}.
\newblock \emph{\bibinfo{journal}{Astron. J.}} \textbf{\bibinfo{volume}{141}},
  \bibinfo{pages}{100} (\bibinfo{year}{2011}).

\bibitem{VanWassenhove:2012:L7}
\bibinfo{author}{Van~Wassenhove, S.} \emph{et~al.}
\newblock {Observability of Dual Active Galactic Nuclei in Merging Galaxies}.
\newblock \emph{\bibinfo{journal}{Astrophys. J. Lett.}}
  \textbf{\bibinfo{volume}{748}}, \bibinfo{pages}{L7} (\bibinfo{year}{2012}).

\bibitem{Springel:2005:1105}
\bibinfo{author}{Springel, V.}
\newblock {The cosmological simulation code GADGET-2}.
\newblock \emph{\bibinfo{journal}{Mon. Not. R. Astron. Soc.}}
  \textbf{\bibinfo{volume}{364}}, \bibinfo{pages}{1105--1134}
  (\bibinfo{year}{2005}).

\bibitem{Hopkins:2007:731}
\bibinfo{author}{Hopkins, P.~F.}, \bibinfo{author}{Richards, G.~T.} \&
  \bibinfo{author}{Hernquist, L.}
\newblock {An Observational Determination of the Bolometric Quasar Luminosity
  Function}.
\newblock \emph{\bibinfo{journal}{Astrophys. J.}}
  \textbf{\bibinfo{volume}{654}}, \bibinfo{pages}{731--753}
  (\bibinfo{year}{2007}).

\bibitem{Hunt:2004:707}
\bibinfo{author}{Hunt, L.~K.} \& \bibinfo{author}{Malkan, M.~A.}
\newblock {Circumnuclear Structure and Black Hole Fueling: Hubble Space
  Telescope NICMOS Imaging of 250 Active and Normal Galaxies}.
\newblock \emph{\bibinfo{journal}{Astrophys. J.}}
  \textbf{\bibinfo{volume}{616}}, \bibinfo{pages}{707--729}
  (\bibinfo{year}{2004}).

\bibitem{Capelo:2015:2123}
\bibinfo{author}{Capelo, P.~R.} \emph{et~al.}
\newblock {Growth and activity of black holes in galaxy mergers with varying
  mass ratios}.
\newblock \emph{\bibinfo{journal}{Mon. Not. R. Astron. Soc.}}
  \textbf{\bibinfo{volume}{447}}, \bibinfo{pages}{2123--2143}
  (\bibinfo{year}{2015}).

\bibitem{Verbiest:2016:1267}
\bibinfo{author}{Verbiest, J. P.~W.} \emph{et~al.}
\newblock {The International Pulsar Timing Array: First data release}.
\newblock \emph{\bibinfo{journal}{Mon. Not. R. Astron. Soc.}}
  \textbf{\bibinfo{volume}{458}}, \bibinfo{pages}{1267--1288}
  (\bibinfo{year}{2016}).

\bibitem{Tang:2018:2249}
\bibinfo{author}{Tang, Y.}, \bibinfo{author}{Haiman, Z.} \&
  \bibinfo{author}{MacFadyen, A.}
\newblock {The late inspiral of supermassive black hole binaries with
  circumbinary gas discs in the LISA band}.
\newblock \emph{\bibinfo{journal}{Mon. Not. R. Astron. Soc.}}
  \textbf{\bibinfo{volume}{476}}, \bibinfo{pages}{2249--2257}
  (\bibinfo{year}{2018}).

\bibitem{Sesana:2018:42}
\bibinfo{author}{Sesana, A.}, \bibinfo{author}{Haiman, Z.},
  \bibinfo{author}{Kocsis, B.} \& \bibinfo{author}{Kelley, L.~Z.}
\newblock {Testing the Binary Hypothesis: Pulsar Timing Constraints on
  Supermassive Black Hole Binary Candidates}.
\newblock \emph{\bibinfo{journal}{Astrophys. J.}}
  \textbf{\bibinfo{volume}{856}}, \bibinfo{pages}{42} (\bibinfo{year}{2018}).

\bibitem{Mayer:2013:244008}
\bibinfo{author}{Mayer, L.}
\newblock {Massive black hole binaries in gas-rich galaxy mergers; multiple
  regimes of orbital decay and interplay with gas inflows}.
\newblock \emph{\bibinfo{journal}{Classical Quant. Grav.}}
  \textbf{\bibinfo{volume}{30}}, \bibinfo{pages}{244008}
  (\bibinfo{year}{2013}).

\bibitem{Lang:2009:094035}
\bibinfo{author}{Lang, R.~N.} \& \bibinfo{author}{Hughes, S.~A.}
\newblock {Advanced localization of massive black hole coalescences with LISA}.
\newblock \emph{\bibinfo{journal}{Classical Quant. Grav.}}
  \textbf{\bibinfo{volume}{26}}, \bibinfo{pages}{094035}
  (\bibinfo{year}{2009}).

\bibitem{Massaro:2009:691}
\bibinfo{author}{Massaro, E.} \emph{et~al.}
\newblock {Roma-BZCAT: a multifrequency catalogue of blazars}.
\newblock \emph{\bibinfo{journal}{Astron. Astrophys.}}
  \textbf{\bibinfo{volume}{495}}, \bibinfo{pages}{691--696}
  (\bibinfo{year}{2009}).

\bibitem{Bertin:1996:393}
\bibinfo{author}{Bertin, E.} \& \bibinfo{author}{Arnouts, S.}
\newblock {SExtractor: Software for source extraction.}
\newblock \emph{\bibinfo{journal}{Astron. Astrophys. Suppl. Ser.}}
  \textbf{\bibinfo{volume}{117}}, \bibinfo{pages}{393} (\bibinfo{year}{1996}).

\bibitem{Blanton:2011:31}
\bibinfo{author}{Blanton, M.~R.}, \bibinfo{author}{Kazin, E.},
  \bibinfo{author}{Muna, D.}, \bibinfo{author}{Weaver, B.~A.} \&
  \bibinfo{author}{Price-Whelan, A.}
\newblock {Improved Background Subtraction for the Sloan Digital Sky Survey
  Images}.
\newblock \emph{\bibinfo{journal}{Astron. J.}} \textbf{\bibinfo{volume}{142}},
  \bibinfo{pages}{31} (\bibinfo{year}{2011}).

\bibitem{deVaucouleurs:1995}
\bibinfo{author}{de~Vaucouleurs, G.} \emph{et~al.}
\newblock {Third Reference Cat. of Bright Galaxies (RC3) (de Vaucouleurs+
  1991)}.
\newblock \emph{\bibinfo{journal}{VizieR On-line Data Catalog: VII/155.
  Originally published in: Springer-Verlag: New York}}
  \textbf{\bibinfo{volume}{7155}}, \bibinfo{pages}{0} (\bibinfo{year}{1995}).

\bibitem{Veron-Cetty:2010:A10}
\bibinfo{author}{Veron-Cetty, M.~P.} \& \bibinfo{author}{Veron, P.}
\newblock {A catalogue of quasars and active nuclei: 13th edition}.
\newblock \emph{\bibinfo{journal}{Astron. Astrophys.}}
  \textbf{\bibinfo{volume}{518}}, \bibinfo{pages}{A10} (\bibinfo{year}{2010}).

\bibitem{Patton:9:235}
\bibinfo{author}{Patton, D.} \& \bibinfo{author}{Atfield, J.}
\newblock {The Luminosity Dependence of the Galaxy Merger Rate}.
\newblock \emph{\bibinfo{journal}{Astrophys. J.}}
  \textbf{\bibinfo{volume}{685}}, \bibinfo{pages}{235--235--246--246}
  (\bibinfo{year}{9}).

\bibitem{Weigel:2018:2308}
\bibinfo{author}{Weigel, A.~K.}, \bibinfo{author}{Schawinski, K.},
  \bibinfo{author}{Treister, E.}, \bibinfo{author}{Trakhtenbrot, B.} \&
  \bibinfo{author}{Sanders, D.~B.}
\newblock {The fraction of AGNs in major merger galaxies and its luminosity
  dependence}.
\newblock \emph{\bibinfo{journal}{Mon. Not. R. Astron. Soc.}}
  \textbf{\bibinfo{volume}{476}}, \bibinfo{pages}{2308--2317}
  (\bibinfo{year}{2018}).

\bibitem{Davies:2015:127}
\bibinfo{author}{Davies, R.~I.} \emph{et~al.}
\newblock {Insights on the Dusty Torus and Neutral Torus from Optical and X-Ray
  Obscuration in a Complete Volume Limited Hard X-Ray AGN Sample}.
\newblock \emph{\bibinfo{journal}{Astrophys. J.}}
  \textbf{\bibinfo{volume}{806}}, \bibinfo{pages}{127} (\bibinfo{year}{2015}).

\bibitem{Koss:2011:57}
\bibinfo{author}{Koss, M.} \emph{et~al.}
\newblock {HOST GALAXY PROPERTIES OF THE SWIFTBAT ULTRA HARD X-RAY SELECTED
  ACTIVE GALACTIC NUCLEUS}.
\newblock \emph{\bibinfo{journal}{Astrophys. J.}}
  \textbf{\bibinfo{volume}{739}}, \bibinfo{pages}{57} (\bibinfo{year}{2011}).

\bibitem{Abazajian:2009:543}
\bibinfo{author}{Abazajian, K.~N.} \emph{et~al.}
\newblock {The Seventh Data Release of the Sloan Digital Sky Survey}.
\newblock \emph{\bibinfo{journal}{Astrophys. J. Suppl. Ser.}}
  \textbf{\bibinfo{volume}{182}}, \bibinfo{pages}{543--558}
  (\bibinfo{year}{2009}).

\bibitem{Blanton:2005:143}
\bibinfo{author}{Blanton, M.~R.}, \bibinfo{author}{Eisenstein, D.},
  \bibinfo{author}{Hogg, D.~W.}, \bibinfo{author}{Schlegel, D.~J.} \&
  \bibinfo{author}{Brinkmann, J.}
\newblock {Relationship between Environment and the Broadband Optical
  Properties of Galaxies in the Sloan Digital Sky Survey}.
\newblock \emph{\bibinfo{journal}{Astrophys. J.}}
  \textbf{\bibinfo{volume}{629}}, \bibinfo{pages}{143} (\bibinfo{year}{2005}).

\bibitem{Kauffmann:2003:33}
\bibinfo{author}{Kauffmann, G.} \emph{et~al.}
\newblock {Stellar masses and star formation histories for 105 galaxies from
  the Sloan Digital Sky Survey}.
\newblock \emph{\bibinfo{journal}{Mon. Not. R. Astron. Soc.}}
  \textbf{\bibinfo{volume}{341}}, \bibinfo{pages}{33} (\bibinfo{year}{2003}).

\bibitem{Brinchmann:2004:1151}
\bibinfo{author}{Brinchmann, J.} \emph{et~al.}
\newblock {The physical properties of star-forming galaxies in the low-redshift
  Universe}.
\newblock \emph{\bibinfo{journal}{Mon. Not. R. Astron. Soc.}}
  \textbf{\bibinfo{volume}{351}}, \bibinfo{pages}{1151} (\bibinfo{year}{2004}).

\bibitem{Chary:2001:562}
\bibinfo{author}{Chary, R.} \& \bibinfo{author}{Elbaz, D.}
\newblock {Interpreting the Cosmic Infrared Background: Constraints on the
  Evolution of the Dust-enshrouded Star Formation Rate}.
\newblock \emph{\bibinfo{journal}{Astrophys. J.}}
  \textbf{\bibinfo{volume}{556}}, \bibinfo{pages}{562--581}
  (\bibinfo{year}{2001}).

\bibitem{U:2012:9}
\bibinfo{author}{U, V.} \emph{et~al.}
\newblock {Spectral Energy Distributions of Local Luminous and Ultraluminous
  Infrared Galaxies}.
\newblock \emph{\bibinfo{journal}{Astrophys. J. Suppl. Ser.}}
  \textbf{\bibinfo{volume}{203}}, \bibinfo{pages}{9} (\bibinfo{year}{2012}).

\bibitem{DasGupta:2001:101}
\bibinfo{author}{DasGupta, A.}, \bibinfo{author}{Cai, T.~T.} \&
  \bibinfo{author}{Brown, L.~D.}
\newblock {Interval Estimation for a Binomial Proportion}.
\newblock \emph{\bibinfo{journal}{Stat. Sci.}} \textbf{\bibinfo{volume}{16}},
  \bibinfo{pages}{101--133} (\bibinfo{year}{2001}).

\bibitem{Hung:2014:63}
\bibinfo{author}{Hung, C.-L.} \emph{et~al.}
\newblock {A Comparison of the Morphological Properties between Local and z
  {\textasciitilde} 1 Infrared Luminous Galaxies: Are Local and High-z (U)LIRGs
  Different?}
\newblock \emph{\bibinfo{journal}{Astrophys. J.}}
  \textbf{\bibinfo{volume}{791}}, \bibinfo{pages}{63} (\bibinfo{year}{2014}).

\bibitem{Grogin:2011:35}
\bibinfo{author}{Grogin, N.~A.} \emph{et~al.}
\newblock {CANDELS: The Cosmic Assembly Near-infrared Deep Extragalactic Legacy
  Survey}.
\newblock \emph{\bibinfo{journal}{Astrophys. J. Suppl. Ser.}}
  \textbf{\bibinfo{volume}{197}}, \bibinfo{pages}{35} (\bibinfo{year}{2011}).

\bibitem{Springel:2003:289}
\bibinfo{author}{Springel, V.} \& \bibinfo{author}{Hernquist, L.}
\newblock {Cosmological smoothed particle hydrodynamics simulations: a hybrid
  multiphase model for star formation}.
\newblock \emph{\bibinfo{journal}{Mon. Not. R. Astron. Soc.}}
  \textbf{\bibinfo{volume}{339}}, \bibinfo{pages}{289--311}
  (\bibinfo{year}{2003}).

\bibitem{Narayan:2008:733}
\bibinfo{author}{Narayan, R.} \& \bibinfo{author}{McClintock, J.~E.}
\newblock {Advection-dominated accretion and the black hole event horizon}.
\newblock \emph{\bibinfo{journal}{New Astron. Rev.}}
  \textbf{\bibinfo{volume}{51}}, \bibinfo{pages}{733--751}
  (\bibinfo{year}{2008}).

\bibitem{Kormendy:2013:511}
\bibinfo{author}{Kormendy, J.} \& \bibinfo{author}{Ho, L.~C.}
\newblock {Coevolution (Or Not) of Supermassive Black Holes and Host Galaxies}.
\newblock \emph{\bibinfo{journal}{Annu. Rev. Astron. Astrophys.}}
  \textbf{\bibinfo{volume}{51}}, \bibinfo{pages}{511--653}
  (\bibinfo{year}{2013}).

\bibitem{Jonsson:2006:2}
\bibinfo{author}{Jonsson, P.}
\newblock {SUNRISE: polychromatic dust radiative transfer in arbitrary
  geometries}.
\newblock \emph{\bibinfo{journal}{Mon. Not. R. Astron. Soc.}}
  \textbf{\bibinfo{volume}{372}}, \bibinfo{pages}{2--20}
  (\bibinfo{year}{2006}).

\bibitem{Jonsson:2010:17}
\bibinfo{author}{Jonsson, P.}, \bibinfo{author}{Groves, B.~A.} \&
  \bibinfo{author}{Cox, T.~J.}
\newblock {High-resolution panchromatic spectral models of galaxies including
  photoionization and dust}.
\newblock \emph{\bibinfo{journal}{Mon. Not. R. Astron. Soc.}}
  \textbf{\bibinfo{volume}{403}}, \bibinfo{pages}{17--44}
  (\bibinfo{year}{2010}).

\bibitem{Narayanan:2010:1701}
\bibinfo{author}{Narayanan, D.} \emph{et~al.}
\newblock {A physical model for z {\textasciitilde} 2 dust-obscured galaxies}.
\newblock \emph{\bibinfo{journal}{Mon. Not. R. Astron. Soc.}}
  \textbf{\bibinfo{volume}{407}}, \bibinfo{pages}{1701--1720}
  (\bibinfo{year}{2010}).

\bibitem{Snyder:2013:168}
\bibinfo{author}{Snyder, G.~F.} \emph{et~al.}
\newblock {Modeling Mid-infrared Diagnostics of Obscured Quasars and
  Starbursts}.
\newblock \emph{\bibinfo{journal}{Astrophys. J.}}
  \textbf{\bibinfo{volume}{768}}, \bibinfo{pages}{168} (\bibinfo{year}{2013}).

\bibitem{Blecha:2013:1341}
\bibinfo{author}{Blecha, L.}, \bibinfo{author}{Civano, F.},
  \bibinfo{author}{Elvis, M.} \& \bibinfo{author}{Loeb, A.}
\newblock {Constraints on the nature of CID-42: recoil kick or supermassive
  black hole pair?}
\newblock \emph{\bibinfo{journal}{Mon. Not. R. Astron. Soc.}}
  \textbf{\bibinfo{volume}{428}}, \bibinfo{pages}{1341--1350}
  (\bibinfo{year}{2013}).

\bibitem{Leitherer:1999:3}
\bibinfo{author}{Leitherer, C.} \emph{et~al.}
\newblock {Starburst99: Synthesis Models for Galaxies with Active Star
  Formation}.
\newblock \emph{\bibinfo{journal}{Astrophys. J. Suppl. Ser.}}
  \textbf{\bibinfo{volume}{123}}, \bibinfo{pages}{3--40}
  (\bibinfo{year}{1999}).

\bibitem{Groves:2008:438}
\bibinfo{author}{Groves, B.} \emph{et~al.}
\newblock {Modeling the Pan-Spectral Energy Distribution of Starburst Galaxies.
  IV. The Controlling Parameters of the Starburst SED}.
\newblock \emph{\bibinfo{journal}{Astrophys. J. Suppl. Ser.}}
  \textbf{\bibinfo{volume}{176}}, \bibinfo{pages}{438--456}
  (\bibinfo{year}{2008}).

\end{thebibliography}
\end{document}